\documentclass[twocolumn]{aastex631}
\revised{July 20, 2021}

\submitjournal{ApJ}

\begin{document}

\title{Luminous Late-time Radio Emission from Supernovae Interacting with Circumbinary Material}

\author[0000-0003-2872-5153]{Samantha C. Wu}
\affiliation{The Observatories of the Carnegie Institution for Science, Pasadena, CA 91101, USA}
\affiliation{Center for Interdisciplinary Exploration \& Research in Astrophysics (CIERA), Physics \& Astronomy, Northwestern University, Evanston, IL 60202, USA}

\author[0000-0002-6347-3089]{Daichi Tsuna}
\affiliation{TAPIR, Mailcode 350-17, California Institute of Technology, Pasadena, CA 91125, USA}
\affiliation{Research Center for the Early Universe (RESCEU), School of Science, The University of Tokyo,  Bunkyo-ku, Tokyo 113-0033, Japan}

\begin{abstract}
Numerous core-collapse supernovae (CCSNe) exhibit signatures of interaction with circumstellar material (CSM). Bright radio emission years after the SN is one such indication of dense CSM at large distances from the star, which may be generated via binary interactions. 
In this work, we use forward modeling to study the radio emission produced by interaction between the SN ejecta and CSM formed by non-conservative stable mass transfer from stripped-envelope stars in short-period binaries. The donors are among the likely progenitors of hydrogen-poor CCSNe that significantly expand $10^3$--$10^4$ years before core-collapse, with companions that best represent low-mass compact objects.
We identify that non-conservative stable mass transfer from lower-mass stripped stars can create a detached shell-like CSM, whereas for our higher-mass stars the CSM is wind-like. In our models, mass transfer rates of $\sim 10^{-4}~M_\odot$ yr$^{-1}$ lead to dense CSM extending to $\sim 10^{18}$~cm. The predicted radio emission is luminous at late times, reaching $L_{\nu}\sim10^{26}$--$10^{29}~\mathrm{erg}~\mathrm{s}^{-1}~\mathrm{Hz}^{-1}$ at years to decades after core-collapse, which is as bright as late-time radio emission observed for a sample of hydrogen-poor SNe. However, the light curves of events with early-time data show more complex behavior in the weeks to months after core-collapse.
We qualitatively demonstrate that similar early-time emission can manifest for CSM that is accelerated to speeds of $\sim10^3~\mathrm{km}~\mathrm{s}^{-1}$ upon ejection, as well as for different viewing angles in case of an asymmetric CSM distribution.

\end{abstract}

\section{Introduction} \label{sec:intro}

Among the population of known core-collapse supernova (SN) explosions, a subset exhibit signatures of interaction between the SN ejecta and circumstellar material (CSM). Interacting SNe are associated with diverse spectroscopic classifications, from hydrogen-rich Type II and Type IIn, to hydrogen-poor Type Ib/c and Type Ibn/Icn SNe (e.g. \citealt{Schlegel90,Pastorello08,2014ApJ...788L..14S,Morozova2017,forster2018,pooley2019,Galyam22,pellegrino2022,bruch2021,bruch2023,jacobsongalan2024}). In these events, interaction is inferred to power the light curves at early or late times and, in some cases, give rise to narrow emission lines.

Although observations of interacting SNe have accumulated evidence that the CSM is generated via relatively high rates of mass loss in a fraction of core-collapse SN progenitors, the mechanisms underlying CSM production remain an open question. Studies of energy injection within the progenitor demonstrate that an input of super-Eddington flux can eject dense CSM \citep[][]{tsang2022,tsuna2023b}, but the source of this energy is as of yet unclear. Proposed explanations include wave-driven mass loss \citep[][]{Quataert2012,Fuller2017,Fuller2018,Wu21,wu2022a}, unstable or explosive burning \citep{Meakin2006,Meakin2007a,Arnett2011,Woosley2015}, or pulsational pair instability \citep{moriya2015}. Another promising explanation is mass loss from interacting binary systems \citep[e.g.,][]{chevalier12_CE,smith2017,dessart2022,wu2022,dong2024,matsuoka2024,ercolino2024}.

Observational indications of dense CSM around some Type Ib SNe include detections of bright radio emission at late times \citep{Stroh21,Rose24}. Such luminous late-time radio emission may be interpreted as interaction between the SN shock and dense CSM formed via large mass loss rates. 
In the framework of radio emission powered by synchrotron emission from electrons accelerated in collisionless shocks \citep[e.g.,][]{Chevalier98,Chevalier06}, the timing of the inferred interaction yields the location of the CSM interacting with the shock. For a typically assumed shock velocity of $\sim 10^4~\mathrm{km}\ \mathrm{s}^{-1}$, radio emission detected at a few--tens of years after core collapse probes material at radii of $10^{17}$--$10^{18}$ cm. 
In order to reach such large radii, dense CSM must be ejected by $10^3$--$10^5$ yr before core collapse, for ejection velocities of $\sim 10$--$10^3~\mathrm{km}\ \mathrm{s}^{-1}$. 

For massive stellar progenitors of core-collapse SNe, these timescales are reminiscent of the nuclear timescale for carbon burning, which has been associated with stellar expansion in studies of core-collapse SN progenitors whose hydrogen-rich envelopes have been stripped by prior binary interaction \citep[][]{dewi2003,habets1986,laplace2020}. In turn, this stellar expansion can initiate Roche lobe overflow for progenitors in close binary systems, thereby producing CSM from binary interaction on timescales relevant to those inferred from the late-time radio emission \citep{tauris2017,wu2022}.
Once Roche lobe overflow occurs, the progenitor and its companion will interact, for instance in the forms of stable mass transfer, common envelope, or stellar merger, each of which processes can produce high mass loss rates that lead to dense CSM.

In this work, we employ a novel forward modeling approach to investigate how CSM produced via non-conservative stable mass transfer can be observed in radio emission. We use binary stellar evolutionary models to simulate the mass loss from donors that represent progenitors of Type Ib SNe residing in short-period binaries, and thereby ascertain the properties of the CSM in such systems. 
Given the CSM density profiles that arise from the binary evolution models, we make predictions for the radio emission from interaction between the SN ejecta and CSM. 
We compare our radio light curves to observed emission for a sample of hydrogen-poor SNe with late-time radio detections, and where possible we also evaluate our models against early-time radio data of these events.

With these methods, we demonstrate that non-conservative stable mass transfer from SN Ib progenitors in short-period binaries onto a low-mass compact companion produces dense CSM, which gives rise to extremely luminous emission at years to decades after core collapse---bright enough to explain observed late-time radio emission from several SNe. Early-time radio data favor the presence of lower-density material in the inner region. Within our framework, we can produce earlier emission with a large CSM velocity of $\sim 10^3$~km~s$^{-1}$, as well as by considering a separate component of the stellar wind complementary to an asymmetric distribution of dense CSM.

In Section \ref{sec:binarymodels}, we describe the binary evolution models we use to produce our CSM profiles. In Sections \ref{sec:dynamics} and \ref{sec:synchrotron}, we detail the framework used to calculate the observed spectrum of synchrotron radio emission, given a model for the mass loss rate and ejection velocity of the CSM.  We present our predicted radio light curves 
in Section \ref{sec:results}, discuss uncertainties in Section \ref{sec:discussion}, and conclude in Section \ref{sec:conclusions}.

\begin{table*}
\centering
\hspace{-2cm} 
\begin{tabular}{cccccc}
\hline
Label & $M_i$ $(M_{\odot})$ & $M_{\rm ej}$ $(M_{\odot})$ & $P_{\mathrm{orb},i}$ (d) & $M_{\rm CSM}$  $(M_{\odot})$& Initial H (Y/N) \\
\hline
$M_{\rm i}=2.9\,M_{\odot}$ (H-free) & 2.90 & 1.2 & 10 & 0.33 & N \\
$M_{\rm i}=3.68\,M_{\odot}$ (H-free) & 3.68 & 2.1 & 1 & 0.18 & N \\
$M_{\rm i}=4.08\,M_{\odot}$ (H-free) & 4.08 & 2.6 & 1 & 0.036 & N \\
$M_{\rm i}=3.0\,M_{\odot}$ (H-poor) & 3.00 & 1.26 & 10 & 0.32 & Y \\
$M_{\rm i}=3.81\, M_{\odot}$ (H-poor) & 3.81 & 2.1 & 1.25 & 0.34 & Y \\
$M_{\rm i}=4.25\,M_{\odot}$ (H-poor) & 4.25 & 2.6 & 0.9 & 0.21 &  Y \\
\hline \\[-1mm]
\end{tabular}
\caption{Properties of the stripped star models presented in this work. Listed are the label used to refer to the model, initial mass of the stripped star $M_i$, estimated ejecta mass at core collapse $M_{\rm ej}$, initial orbital period of the binary $P_{\mathrm{orb},i}$, amount of CSM mass lost through mass transfer $M_{\rm CSM}$, and whether or not the stripped star initially retained a hydrogen (H) envelope at the onset of the binary simulation. }
\label{tab:hestarmodels}
\end{table*}

\section{Radio emission from CSM produced by binary interaction} \label{sec:methods}

We analyze the evolution of SN ejecta with mass $M_{\rm ej}$ and energy $E_{\rm ej}$ sweeping up CSM whose density is given by a profile $\rho_{\rm CSM}(r)$. The progenitor systems that produce the CSM are modeled as binaries consisting of a stripped-envelope star (or stripped star) with a companion $M_c =1.4\, M_{\odot}$, in which expansion of the stripped star during late-stage nuclear burning leads to non-conservative mass transfer onto the companion. We consider the mass to leave the system in an equatorial outflow and form a circumbinary torus of CSM \citep{pejcha2016,MacLeod2018}. 

\subsection{Binary evolution models}
\label{sec:binarymodels}

In this section, we describe the details of our stellar evolution models for producing CSM from binary interaction. We model the binary evolution of stripped stars at $Z = 0.02$ with a $M_{\rm c} = 1.4\, M_{\odot}$ companion in MESA \citep[version r15140,][]{mesa2011,mesa2013,mesa2015,mesa2018,mesa2019}. The $M_c =1.4\, M_{\odot}$ companion is represented by a point mass in the simulation. Throughout the evolution, we assume a circular orbit and fully non-conservative mass transfer (e.g. $f_{\rm mt}=0,\ \beta_{\rm mt}=1$ as in \citealt{mesa2015}) where mass and angular momentum are removed from the system in the vicinity of the accretor as a fast wind. We follow the stellar evolution of the stripped star from core helium (He) burning up to at least oxygen (O) burning, and where possible until silicon burning.  
Our methods to create and evolve the stripped stars follow those of \cite{wu2022}.\footnote{Inlists used available on Zenodo under an open-source 
Creative Commons Attribution license: \dataset[doi:10.5281/zenodo.7106182]{https://doi.org/10.5281/zenodo.7106182}}
As in that work, we use a modified version of the implicit mass transfer scheme of \citet{kolb1990} for Roche lobe overflow that is revised to account for both radiation and gas pressure \citep[e.g.,][]{marchant2021}. 

The initial stripped-star masses $M_{i}$ and orbital periods $P_{\mathrm{orb},i}$ of our models, along with their labels, ejecta masses, and CSM masses, are listed in Table \ref{tab:hestarmodels}. 
For a given metallicity and mass ratio, the initial mass of the stripped star largely determines its radius evolution. Higher-mass stripped stars expand monotonically and achieve smaller stellar radii of only $R_*\lesssim$ a few $R_\odot$, while lower-mass stripped stars of $M_i \lesssim 3\, M_{\odot}$ exhibit non-monotonic radius evolution and expand to much larger radii of $R_*\approx 10$--$100~R_\odot$ \citep{laplace2020,wu2022}. 

We expect the radius evolution characteristic of higher masses to lead to a continuous, wind-like CSM profile when these stars fill their Roche lobes, which is possible at short orbital periods on the order of days. Meanwhile, lower-mass stripped stars are more likely to exhibit detached shells in their CSM profiles, since they can detach from their Roche lobes when they contract during the late stages of their evolution \citep{wu2022}. The initial binary periods of our models are chosen such that the donors will undergo high rates of mass transfer from $\sim 10^4$ yr before core collapse. The limited expansion of stripped stars with $M_i \gtrsim 4.5\, M_{\odot}$ for solar metallicity motivates the upper limit of $M_{i}$ explored in this work \citep{laplace2020}, whereas the expected ejecta mass for typical stripped-envelope SNe ($M_{\rm ej}=1$--$3~M_\odot$) precludes our inclusion of lower-mass stars \citep{Drout11,Lyman16,Taddia18}. 

In some of our models, we remove the entire hydrogen (H) envelope before the onset of core He burning, whereas in others, we allow the star to retain some H at the onset of our binary evolution simulations (noted in Table \ref{tab:hestarmodels}). 
At the onset of core carbon (C) burning, the H-poor models retain a total mass of $\sim 10^{-2}\, M_{\odot}$ of H, which is consistent with the predicted H mass at the end of core He burning for solar metallicity stars from, e.g., \cite{laplace2020}. In principle, the amount of H-rich envelope at the onset of core C burning is sensitive to the binary evolution history of the stripped star, as well as modeling choices when creating the stripped star model. Variations in the H mass retained after He burning will alter the composition and density of the outermost CSM in these models.

We assume non-conservative mass transfer where the mass is removed from the system in the vicinity of the accretor. 
The assumption that nearly all the mass transferred is lost from the system is appropriate in the case of a neutron star (NS) companion, since we find mass transfer rates in our binary models of $\gtrsim 10^{-4}\, M_{\odot}\, \rm{yr}^{-1}$ that exceed the Eddington-limited rate of a NS ($\dot{M}_{\rm Edd} \sim 4\times 10^{-8}\, M_{\odot}\, \rm{yr}^{-1}$) by many orders of magnitude. As the companion is represented as a point mass in our binary simulations, our calculations are most representative of a scenario where the accretor is a $M_c=1.4\, M_{\odot}$ NS companion.

\begin{figure*}
    \centering
    \includegraphics[width=0.495\textwidth]{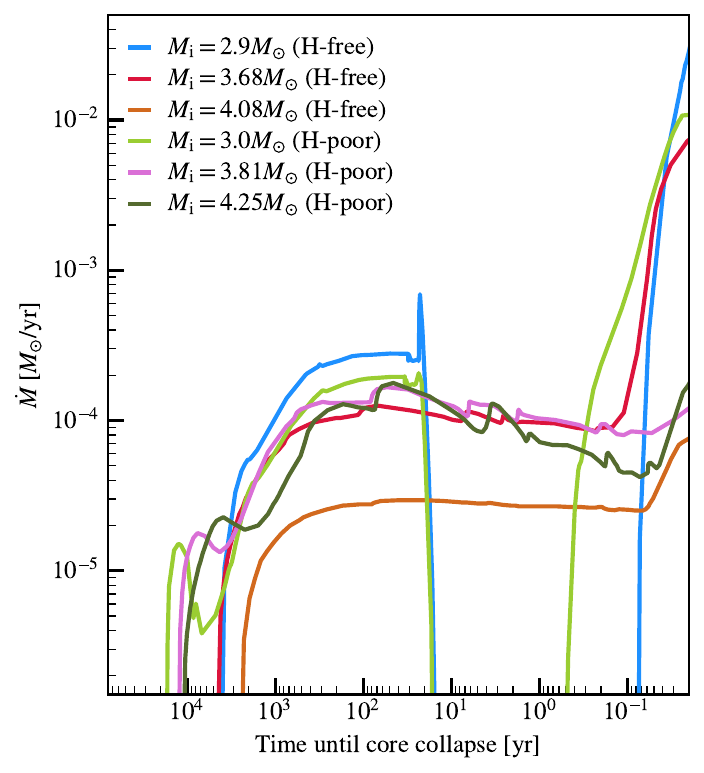}
    \includegraphics[width=0.495\textwidth]{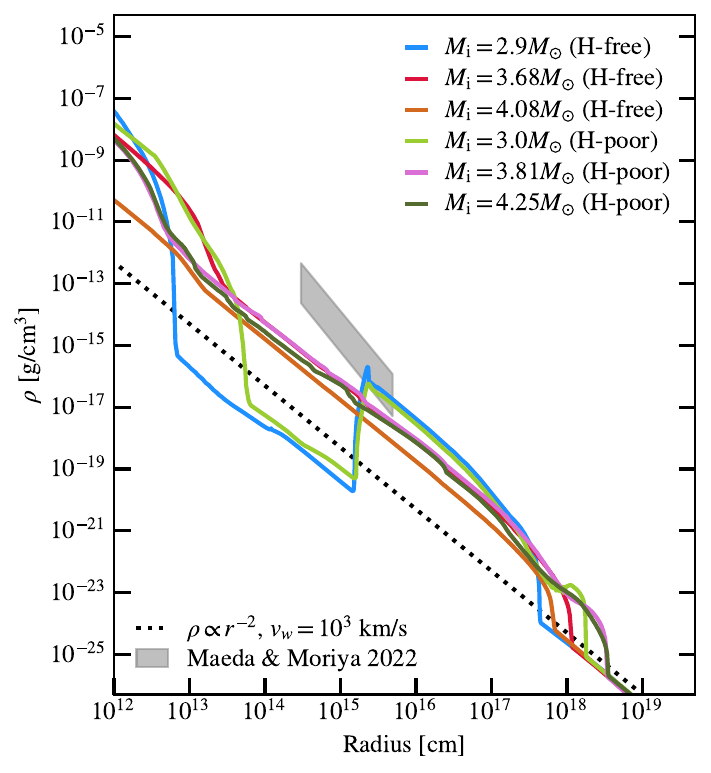}
    \caption{\textit{Left}: The mass loss history of the stripped star models listed in Table \ref{tab:hestarmodels}, shown as the mass transfer rate of the donor star as a function of time until core collapse. \textit{Right}: Density profiles of the CSM for the stripped star models listed in Table \ref{tab:hestarmodels}, shown for parameters $v_{\rm CSM} = 0.3\, v_{\mathrm{orb},c}$ and $f_{\Omega}=1$. The dotted line shows a wind profile CSM for a mass-loss rate $10^{-5}\ M_\odot\ {\rm yr}^{-1}$ and velocity of $10^{3}~\mathrm{km}\ \mathrm{s}^{-1}$, typically observed in Wolf-Rayet stars. The gray shaded region is taken from Figure 7 of \cite{Maeda22} and represents the range of CSM density distributions derived for SNe Ibn in that work.} 
    \label{fig:density}
\end{figure*}

\subsubsection{Mass loss history}
\label{sec:massloss}
In our binary evolution models, the stripped donor star begins mass transfer after core He exhaustion and during C burning. The left panel of Figure \ref{fig:density} shows the mass loss history for our stripped stars. During the first mass transfer phase from $\sim 10^{3}$--$10^{4}$ yr before core collapse, the mass loss rates reach $\sim 10^{-5}$--$10^{-4}\, M_{\odot}\, \rm{yr}^{-1}$. The stripped star models labeled  $M_i = 2.9\, M_{\odot}$ (H-free) and $M_i = 3.0\, M_{\odot}$ (H-poor) exhibit  the most elevated mass loss rates, reaching a few $\times 10^{-4} \, M_{\odot}\, \rm{yr}^{-1}$ before detaching at $\sim 20$ yr before core collapse. These models then overflow their Roche lobes again at $\sim0.08$--$0.5$ yr before core collapse. The $M_i = 2.9\, M_{\odot}$ (H-free) model represents the upper mass range of models explored in \cite{wu2022} and thus exhibits the late-time mass loss studied in that work. For the other four models presented here, the donor does not detach from its Roche lobe before the end of the simulation. 

Mass loss due to stellar winds is not modeled explicitly in our MESA binary evolution models. We incorporate the effect of wind mass loss into the mass loss history of our models via the following equation from \cite{Nugis2000}:
\begin{equation}
    \label{eq:Mdotwind}
   \log \dot{M}_{\rm w} =1.29\log L  + 1.73 \log Y+\, 0.47 \log Z  -11,
\end{equation}
where $L, Y, Z$ are the stellar luminosity, surface helium abundance and surface metallicity, respectively.
This was derived from Wolf-Rayet stars more massive than our stripped stars. The true mass loss rate for stripped stars in the mass range modeled in this work is uncertain: it may be lower given the rates inferred for local lower-mass stripped stars \citep[e.g.,][]{Gotberg23}, but other studies also suggest stronger stripped star winds \citep{Sander20,Moriya22}.
In our models, we typically see $\dot{M}_{\rm w} \approx 10^{-7}$--$10^{-6}\, M_{\odot}\ \mathrm{yr}^{-1}$. 
We assume the wind is spherically symmetric and lost at the escape velocity of the star, $v_{\rm esc} = \sqrt{2GM_*/R_*}$ where $M_*$ is the stellar mass. The ensuing wind density profile is
\begin{equation}\label{eq:winddensity}
    \rho_{\rm w}(r) = \frac{\dot{M}_{\rm w}}{4\pi r^2 v_{\rm esc}}.
\end{equation}

Using our mass loss rates from mass transfer during the binary simulation, $\dot{M}_{\rm CSM}$ and including the contribution from stellar winds, the density profile of the CSM is estimated as:
\begin{equation}\label{eq:density}
    \rho_{\rm CSM}(r) = \frac{\dot{M}_{\rm CSM}}{4\pi r^2 v_{\rm CSM} f_\Omega} + \rho_{\rm w}(r).
\end{equation}
Here, the covering fraction $f_{\Omega}$ parameterizes the asymmetry of the CSM formed from binary interaction, as this CSM may be expected to form a toroidal structure that covers a fraction $f_{\Omega}$ of the full sphere \citep[][]{pejcha2016,MacLeod2018}\footnote{Simulations show spiral structures in the mid-plane that may translate to inhomogeneities in the density profile, but we assume here that they are minor enough that they do not affect the radio light curves \citep[e.g. Fig. 5 of][]{pejcha2016}. }. 
In order to construct the CSM density profile for each of our binary models, we calculate the mass loss rate $\dot{M}_{\rm CSM}$ and ejection velocity $v_{\rm CSM}$ as a function of time.

In our picture for the circumbinary CSM formation, the high mass transfer rates of our models may lead to mass loss through the L2 point, potentially via a geometrically thick accretion disk \citep[e.g.,][]{lu2022}. We therefore expect the velocity of the CSM to be related to the orbital velocity of the NS companion, $v_{\rm orb, c}$. Smoothed-particle hydrodynamical simulations of mass loss from the outer Lagrange point indicate that unbound outflows reach mean asymptotic velocities of a fraction of the binary escape velocity, so that the CSM velocity is approximately 

\begin{eqnarray}
    v_{\rm CSM} &=& f_{\infty}\! \sqrt{2} (1+q) v_{\rm orb,c} = f_\infty\sqrt{2}\left[\frac{2\pi G(M_*+M_c)}{P_{\rm orb}}\right]^{1/3}\nonumber \\ 
    &\sim& 100\ {\rm km\ s^{-1}}\left(\frac{f_\infty}{0.2}\right)\!\left(\frac{M_*+M_c}{5~M_\odot}\right)^{1/3}\! \left(\frac{P_{\rm orb}}{1~{\rm day}}\right)^{-1/3}\!.
\end{eqnarray}
The fraction $f_\infty$ itself also depends on the mass ratio $q=M_{\rm c}/M_{\rm *}<1$ \citep{pejcha2016}. For our systems of interest, 
$q \approx 0.35$--$0.55$ and $f_\infty \approx 0.15$--$0.25$, so we take $v_{\rm CSM} \sim  0.3\, v_{\rm orb, c}$ for our fiducial CSM velocity, resulting in typical values of $v_{\rm CSM} \sim 100\ {\rm km\ s^{-1}}$ ---though we discuss the outcomes for different CSM velocities in Section \ref{sec:fastcsm}. We estimate the distance that the CSM reaches after leaving the binary system as $r_{\rm CSM} = R_{*} + t_{cc}v_{\rm CSM}$, where $t_{cc}$ is the time remaining until core collapse when the mass is lost.

Given $\dot{M}_{\rm CSM}$, $v_{\rm CSM}$, and $r_{\rm CSM}$ as functions of time for each binary evolution model, we calculate the CSM density profile $\rho_{\rm CSM}(r)$ at the time of SN using Equation \ref{eq:density}. The right panel of Figure \ref{fig:density} shows density profiles for parameters $v_{\rm CSM} = 0.3\, v_{\mathrm{orb},c}$ and $f_{\Omega}=1$. Notably, the density profiles of the $M_i = 2.9\, M_{\odot}$ (H-free) model and the $M_i = 3.0\, M_{\odot}$ (H-poor) model rise sharply to form a dense shell of CSM located at $\gtrsim 10^{15}$ cm. This occurs because mass transfer ceases while the star detaches from its Roche lobe during the evolution of this model. During the period of detachment, the CSM density profile is populated by the spherically-symmetric stellar wind ejected at a few~$\times~10^2~\mathrm{km}\ \mathrm{s}^{-1}$ with $\dot{M}_{\rm w} \lesssim 10^{-6}\, M_{\odot}\ \mathrm{yr}^{-1}$. 

The other density profiles follow an overall decline similar to a $\rho \propto r^{-2}$ wind, but with densities higher by orders of magnitude than that produced by typical assumptions of a $10^3~\mathrm{km}\ \mathrm{s}^{-1}$, $\dot{M}=10^{-5}\, M_{\odot}\ \mathrm{yr}^{-1}$ stellar wind, as motivated from Galactic Wolf-Rayet stars (\citealt{crowther2007}, black dotted line). This dense CSM extends to $\sim 10^{18}$ cm from the star, as expected for mass transfer due to stellar expansion initiated by the time of core C burning.  

For the H-poor models, only the CSM exterior to $10^{18}$ cm is H-rich, with H mass fraction of $X_H>0.1$. This H-rich  material is lost in a phase of mass transfer with mass loss rates of $\sim 10^{-5}\, M_{\odot}\, \mathrm{yr}^{-1}$ at $\sim 10^4 $ yr before core collapse. Any mass lost much earlier than $10^4$ yr has had time to expand past $\gtrsim 10^{18}$ cm. In particular, material ejected during the mass loss phase that created the stripped star $\sim 10^6$ yr before core collapse will be too distant to impact the radio light curves on $\sim$ decade timescales of our interest.

Varying the parameters $v_{\rm CSM}$ and $f_{\Omega}$ changes the density profile of the CSM, excluding regions dominated by the spherically-symmetric stellar wind. For smaller values of $f_{\Omega}$, the magnitude of the density profile is higher throughout by a factor of $f_{\Omega}^{-1}$. For larger values of $v_{\rm CSM}$, the density also decreases as the same CSM mass is spread across a larger volume, and features such as the dense shell move out to larger radii.

\subsection{Dynamics of Ejecta-CSM Interaction} \label{sec:dynamics}

In this section, we describe our formalism for modeling the dynamical evolution of an SN ejecta with mass $M_{\rm ej}$ and energy $E_{\rm ej}$ sweeping up CSM with density given by a profile $\rho_{\rm CSM}(r)$.
The density profile of the CSM produced by binary mass loss is given by Equation \ref{eq:density} and calculated from binary evolution models, as described in detail in Section \ref{sec:binarymodels}. 
We can also estimate the ejecta mass $M_{\rm ej} = M_{\rm final} - M_{\rm NS}$, where $M_{\rm final}$ is the final mass of the stripped star at the end of the MESA simulation and $M_{\rm NS}=1.4~M_\odot$ is the mass of the remnant NS.
The MESA models are run until at least late O burning, and the mass at core-collapse is expected to be within $\lesssim 0.01 M_{\odot}$ of $M_{\rm final}$. 
Our stripped star models produce ejecta masses of $M_{\rm ej}=1$--$3~M_\odot$ (Table \ref{tab:hestarmodels}), consistent with values typically inferred in Type Ibc SNe \citep[e.g.,][]{Drout11,Lyman16,Taddia18}.

We solve the shocked region formed by the homologously expanding ejecta and the CSM, assuming the region is a thin shell with radius $r_{\rm sh}$ and velocity $v_{\rm sh}$ \citep[e.g.,][]{Moriya_et_al_13,Murase24}. 
Due to the large range of CSM densities in our model, the shocks can be either radiative where the swept-up material cools within a dynamical time, or adiabatic where the swept-up material does not cool. In order to capture both regimes, we additionally include the evolution of the shell's internal energy $E_{\rm int}$ as it is dissipated by the ejecta-CSM interaction. We make the simplifying assumption that energy dissipation is dominated by the forward shock interacting with the dense CSM, and therefore evolve only the internal energy generated by the forward shock, $E_{\rm int,fs}$.

We evolve the shock properties in time by mass, momentum, and energy conservation. The mass and momentum conservation are given by
\begin{eqnarray}
    \frac{dM_{\rm sh}}{dt} &=& 4\pi r_{\rm sh}^2 f_\Omega [\rho_{\rm ej}(v_{\rm ej}-v_{\rm sh})+\rho_{\rm CSM}(v_{\rm sh}-v_{\rm CSM})] \nonumber \\ \nonumber\\
    M_{\rm sh}\frac{dv_{\rm sh}}{dt} &=& 4\pi r_{\rm sh}^2 f_\Omega[\rho_{\rm ej}(v_{\rm ej}-v_{\rm sh})^2 - \rho_{\rm CSM}(v_{\rm sh}-v_{\rm CSM})^2] \nonumber \\ 
    && + \frac{2E_{\rm int,fs}}{r_{\rm sh}} \label{eq:dvshdt}
\end{eqnarray}
where $v_{\rm ej}=r_{\rm sh}/t$ is the velocity of the ejecta at $r_{\rm sh}$, and the last term in Equation \ref{eq:dvshdt} is the $PdV$ work done when gas pressure $P$ dominates, with $E_{\rm int,fs}$ evolved by Equation \ref{eq:energycons} below. The SN ejecta is assumed to have a density profile of a double power-law \citep{Matzner99}
\begin{eqnarray}
\rho_{\rm ej} (r,t)
&=& \left\{ \begin{array}{ll}
t^{-3}\left[r/(gt)\right]^{-n} & (r/t > \upsilon_t),\\
t^{-3}(\upsilon_t/g)^{-n} \left[r/(t\upsilon_t)\right]^{-\delta}  & (r/t < \upsilon_t)
\end{array}\right.
\label{eq:rho_ej}
\end{eqnarray}
which is valid roughly after the ejecta expands to a few times the stellar radii and kinetic energy dominates over internal energy. The constants $g$ and $\upsilon_t$ are the following functions of the ejecta mass $M_{\rm ej}$ and energy $E_{\rm ej}$:
\begin{eqnarray}\label{eq:coeff_ej}
g &=& \left\{\frac{1}{4\pi(n-\delta)} \frac{[2(5-\delta)(n-5)E_{\rm ej}]^{(n-3)/2}}{[(3-\delta)(n-3)M_{\rm ej}]^{(n-5)/2}}\right\}^{1/n}\\
\upsilon_t &=& \left[\frac{2(5-\delta)(n-5)E_{\rm ej}}{(3-\delta)(n-3)M_{\rm ej}}\right]^{1/2}. \label{eq:v_t}
\end{eqnarray}
We adopt $n\approx 10$, $\delta\approx 1$ as expected for explosions of a star with a radiative envelope \citep{Chevalier89,Matzner99}.

The forward shock internal energy $E_{\rm int,fs}$ evolves as
\begin{eqnarray}
\label{eq:energycons}
        \frac{dE_{\rm int,fs}}{dt} &=& 4\pi r_{\rm sh}^2f_{\Omega}\left[\frac{2}{(\gamma+1)^2}\right]\rho_{\rm CSM}(v_{\rm sh}-v_{\rm CSM})^3 \nonumber \\
        && - \epsilon_{\rm cool}(\rho_{\rm down}, T_{\rm down})\times \frac{\gamma-1}{\gamma+1}\frac{4\pi f_\Omega r_{\rm sh}^3}{3} \nonumber \\
        && - \frac{2E_{\rm int,fs}v_{\rm sh}}{r_{\rm sh}},
\end{eqnarray}
where for an adiabatic index of $\gamma = 5/3$, the downstream density and temperature from the Rankine-Hugoniot jump conditions are $\rho_{\rm down} = 4\rho_{\rm CSM}$ and
\begin{eqnarray}
    T_{\rm down} &=& \frac{3\mu m_p}{16k_B}v_{\rm sh}^2 \nonumber \\
    &\sim& 3\times 10^9\ {\rm K}\left(\frac{\mu}{4/3}\right)\left(\frac{v_{\rm sh}}{10^4\ {\rm km\ s^{-1}}}\right)^2.
\end{eqnarray} 
At such high temperatures, cooling mainly comes from free-free emission of the fully ionized post-shock gas, with emissivity \citep{Radipro}
\begin{eqnarray}
    \label{eq:emissivity}
    \epsilon_{\rm cool} = 1.4\times 10^{-27} \mathrm{erg}\, \mathrm{cm}^{-3}\, \mathrm{s}^{-1}(T_{\rm down}^{1/2}Z^2 n_e n_i\bar{g}_B)
\end{eqnarray}
where $n_e, n_i$ are respectively the number density in ${\rm cm^{-3}}$ of electrons and ions in the CSM, $Z$ specifies the charge of each ion, and $\bar{g}_B\approx 1.2$ is a frequency-averaged Gaunt factor.
We adopt $Z=2$, $n_e=\rho_{\rm down}/(2m_p)$, $n_i=\rho_{\rm down}/(4m_p)$, and $\mu \approx 4/3$ as expected for fully ionized, helium-rich CSM.

Our equations for the shock dynamics simplistically account for the CSM asymmetry through the covering fraction parameter $f_{\Omega}$. We assume that only a fraction $f_\Omega$ of the (spherical) SN ejecta interacts with the CSM, which is concentrated around the binary's equatorial plane for $f_\Omega < 1$ with an enhanced density $\propto f_\Omega^{-1}$ compared to the spherical case (Equation \ref{eq:density}). The other polar region covering $(1-f_\Omega)$ of the ejecta is assumed to expand into vacuum, and we neglect the contribution of radio emission from these regions. This is a reasonable approximation at the timescales of years to decades of our interest for late-time radio emission, as the polar regions are occupied only by the diffuse wind from the stripped star.

\subsection{Particle acceleration and synchrotron emission}
\label{sec:synchrotron}
We consider particle acceleration from collisionless shocks, which are expected to develop when the shock is no longer radiation mediated \citep[e.g.,][]{Levinson20}. This corresponds to when the radiation downstream of the shock begins to escape efficiently, called ``shock breakout" in the context of SNe. We can thus define the onset of the collisionless shock $t_{\rm 0}$ from the condition for shock breakout \citep[e.g.,][]{Murase18}
\begin{eqnarray}
\label{eq:tauesc}
    \int^{\infty}_{r_{\rm 0}} (\kappa\rho_{\rm CSM})dr = \frac{c}{v_{\rm 0}},
\end{eqnarray}
where $c$ is the speed of light, $r_{\rm 0}=r_{\rm sh}(t=t_{\rm 0})$, and $v_{\rm 0}=v_{\rm sh}(t=t_{\rm 0})$. The opacity $\kappa$ in the CSM is typically scattering dominated for SN shocks, such that $\kappa \approx 0.2\ {\rm cm^2\ g^{-1}}$ for an ionized, hydrogen-poor CSM. 

The electrons injected to the shocked region cool by emitting synchrotron radiation and by adiabatic expansion of the shocked region. We follow the evolution of the number of relativistic electrons at a given Lorentz factor, including injection of newly accelerated particles and their cooling. The radio synchrotron emission is obtained using the energy spectra of relativistic electrons. 

We parameterize the magnetic field strength in the shocked region by the efficiency parameter $\epsilon_B$, 
which scales the magnetic energy density by the cumulative energy dissipated by the shock:
\begin{eqnarray}
    &&\frac{B^2}{8\pi}\left(\frac{4\pi r_{\rm sh}^3 f_\Omega}{3}\right) \nonumber\\
    &=& \epsilon_{\rm B}\int^{t}_{t_{\rm 0}} 4\pi r_{\rm sh}(t')^2 f_\Omega \rho_{\rm CSM}(t') v_{\rm sh}(t')^3 \frac{r_{\rm sh}(t')}{r_{\rm sh}(t)} dt' .\label{eq:magnetic_field}
\end{eqnarray}
Here the final factor takes into account the energy loss due to adiabatic expansion. This formalism is more appropriate than using only the local values of $\rho_{\rm CSM}$ and $v_{\rm sh}$ when these values exhibit abrupt changes with radius, for instance when the CSM profile features a detached dense shell of material.

The number density of relativistic electrons injected into the shocked region is assumed to follow a power-law energy distribution in Lorentz factor (or energy), $dn(\gamma_e)/d\gamma_e=n_0 \gamma_e^{-p}$ ($p>2$), as expected for diffusive shock acceleration. The spectral index of the radio emission is correlated with the power-law index $p$. We assume here $p=3$, as found from radio modeling for Type Ib/c SNe \citep[e.g.,][]{Chevalier06,Maeda12}. 

We scale the normalization of the distribution with the parameter $\epsilon_E$, which describes the fraction of the energy density of relativistic electrons compared to the ram pressure: 
\begin{eqnarray}\label{eq:electron_distribution}
    \int_{\gamma_{\rm min}}^{\infty} (\gamma_e m_e c^2)\frac{dn}{d\gamma_e}d\gamma_e =\epsilon_E \rho_{\rm CSM} v_{\rm sh}^2.
\end{eqnarray}
For $p=3$ this leads to a normalization of
\begin{eqnarray}
    n_0 = \frac{(\epsilon_E\gamma_{\rm min})\rho_{\rm CSM}v_{\rm sh}^2}{m_e c^2},
    \label{eq:electron_n0}
\end{eqnarray}
where $m_e$ is the electron mass. 
Following the standard literature, we assume the non-thermal electron spectra extends to $\gamma_{\rm min}=1$ \citep{Chevalier98}. We note that this assumption can be problematic for high shock velocities of $v_{\rm sh}\gtrsim 0.2c$, since in that regime the thermal (Maxwellian) electron population can reach relativistic energies and contribute to the radio emission \citep{Margalit24}. In our models the shock velocities are typically $v_{\rm sh}\approx (0.01$--$0.1)c$ at all times, small enough to avoid such effects.

The non-thermal electron population at each time $t$ is mediated by the injection of electrons and their cooling. The electron number spectrum $dN(\gamma_e)/d\gamma_e$ evolves as\footnote{Note the difference in dimension, where $n$ refers to number density and $N$ to number.}
\begin{eqnarray}
      \frac{dN/d\gamma_e}{dt} = \frac{\partial}{\partial\gamma_e}\left[\frac{dN}{d\gamma_e}\left(\dot{\gamma}_{e,\rm ad}+\dot{\gamma}_{e,\rm rad}\right)\right] + q_e
\end{eqnarray}
where the adiabatic cooling $\dot{\gamma}_{e,\rm ad}$, synchrotron cooling $\dot{\gamma}_{e,\rm rad}$, and electron injection $q_e$ terms are 
\begin{eqnarray}
      \dot{\gamma}_{e,\rm ad} &=& -\frac{v_{\rm sh}}{r_{\rm sh}}\gamma_e  \\
      \dot{\gamma}_{e,\rm rad} &=& -\frac{\sigma_TB^2}{6\pi m_ec}\gamma_e^2 \\
      q_e &=& 4\pi r_{\rm sh}^2f_\Omega v_{\rm sh}\frac{dn}{d\gamma_e}
\end{eqnarray}
and $\sigma_{\rm T}$ is the Thomson cross section. The initial condition at $t=0$ is $dN/d\gamma_e=0$ for all $\gamma_e$.

From the electron spectrum $dN/d\gamma_e$ at a given time, the synchrotron luminosity emitted at this time is
\begin{eqnarray}
    L_{\nu, \rm syn} = \int d\gamma_e \frac{dN}{d\gamma_e} P_\nu(\gamma_e),
\end{eqnarray}
where $P_\nu(\gamma_e)$ is the synchrotron spectrum produced by a single electron \citep[][]{Radipro},
\begin{eqnarray}
    P_\nu(\gamma_e) = \frac{2\sqrt{3}e^3 B}{3m_ec^2}F(\nu/\nu_c).
\end{eqnarray}
Here, $e$ is the electron charge and $\nu_c=eB\gamma_e^2/2\pi m_e c$ is the characteristic frequency of synchrotron radiation from an electron of Lorentz factor $\gamma_e$.
We use an analytic fit to the synchrotron function $F(x)$ \citep{Fouka2013}. 

At low frequencies and high densities, synchrotron self-absorption (SSA) and free-free absorption (FFA) become important. The SSA absorption coefficient is given by \citep{Radipro}
\begin{equation}
    \alpha_{\rm ssa}(\nu)\! \approx -\frac{1}{8\pi\nu^2m_e}\!\int d\gamma_e \gamma_e^2P_\nu(\gamma_e) \frac{\partial}{\partial \gamma_e}\!\left(\frac{1}{\gamma_e^2}\frac{1}{V}\frac{dN}{d\gamma_e}\right)
\end{equation}
where $V=4\pi r^2f_\Omega \Delta r$ is the volume of the shocked region where the relativistic electrons are located. Here, $\Delta r$ is the radial extent of this region, but the absorption is independent of $\Delta r$ as shown in Equation \ref{eq:tau_absorb}.

The ionized CSM ahead of the shock can attenuate radio waves by FFA, with an absorption coefficient \citep{Mezger67}
\begin{eqnarray}
    \alpha_{\rm ff}(\nu)&=&3.8\times 10^{-29}\ {\rm cm^{-1}}\left(n_e\sum n_iZ^2\right)_{\rm CSM}\nonumber \\
    &&\times\left(\frac{T_{\rm e, CSM}}{10^5\ {\rm K}}\right)^{-1.35}\left(\frac{\nu}{{10\ \rm GHz}}\right)^{-2.1},
\end{eqnarray}
where $T_{\rm e, CSM}$ is the electron temperature in the CSM. 
As in Section \ref{sec:dynamics}, we adopt the parameters expected for fully ionized helium-rich gas of $n_e=\rho_{\rm CSM}/2m_p, n_i=\rho_{\rm CSM}/4m_p$, $Z=2$, and we take $T_{\rm e, CSM}\sim 10^5$ K. 

Given the absorption coefficients, the optical depths for SSA and FFA are then 
\begin{equation}
    \tau_{\rm ssa} \approx \alpha_{\rm ssa}(\nu)\Delta r,\ \tau_{\rm ff}=\int_{r_{\rm sh}}^{\infty}\alpha_{\rm ff}(\nu)dr,
    \label{eq:tau_absorb}
\end{equation}
and we obtain the observed spectrum as
\begin{eqnarray}
    L_{\nu, \rm obs} \approx L_{\nu, \rm syn}\exp(-\tau_{\rm ff}) \frac{1-\exp(-{\tau_{\rm ssa}})}{\tau_{\rm ssa}}.
\end{eqnarray}

We note that for an asymmetric CSM of $f_{\Omega} < 1$ the absorption is complicated in reality, as it depends on the viewing angle and the ionization of the ejecta when viewed from polar angles. If the ejecta is neutral, our estimates in Equation \ref{eq:tau_absorb} represent maximal absorption when viewed along the direction of the CSM, and hence our radio light curves are conservative estimates well before peak. Nevertheless, the ejecta of SNe Ibc are expected to be partially ionized at early times \citep[e.g.,][Figure 13]{Dessart15}, which can effectively mask the radio emission from the CSM torus when viewed from polar angles. The absorption also depends on the uncertain electron temperature in the CSM, $T_{e, \rm CSM}$. These affect our predictions for the early emission, but we verify that at late times ($\gtrsim 1$ yr) the free-free optical depth drops well below unity, so the late-time light curve depends neither on $T_{e, \rm CSM}$ nor the viewing angle.

\section{Radio light curves}
\label{sec:results}

Throughout our forward modeling, we use a fiducial ejecta kinetic energy of $E_{\rm ej} = 10^{51}$ erg to generate our radio light curves, and consider a range of $E_{\rm ej}=(0.5$--$1.5)\times 10^{51}$ erg typically found in Type Ibc SNe \citep{Lyman16,Taddia18}. We vary the parameters $f_{\Omega}$, $\epsilon_B$, $\epsilon_E$, as labeled on the top right of each figure panel. We adopt the microphysical parameters of $(\epsilon_{\rm B}, \epsilon_{E})=10^{-2}$--$10^{-1}$, motivated from modeling of radio supernovae \citep[e.g.,][]{Chevalier06,bietenholz2021}\footnote{ While our values of $\epsilon_E\sim 10^{-2}$--$10^{-1}$ differ at face value from choices made in some other studies, such as $\epsilon_E\sim 10^{-4}$--$10^{-3}$ in \cite{Murase19}, what sets the energy spectra of non-thermal electrons is the product $\epsilon_E\gamma_{\rm min}$ (for $p=3$; Equation \ref{eq:electron_n0}). \cite{Murase19} adopt a larger $\gamma_{\rm min}\sim (m_p/m_e)(v_{\rm sh}/c)\sim 60(v_{\rm sh}/10^4\ {\rm km\ s^{-1}})$ than our formalism, which fixes $\gamma_{\rm min}$ as $1$, leading to values of $\epsilon_E\gamma_{\rm min}$ comparable to ours.}. Where not stated, we assume a CSM ejection speed of $v_{\rm CSM} = 0.3\, v_{\mathrm{orb},c}$, based on the framework outlined in Section \ref{sec:massloss}.

\begin{figure}
    \centering
    \includegraphics[width=\columnwidth]{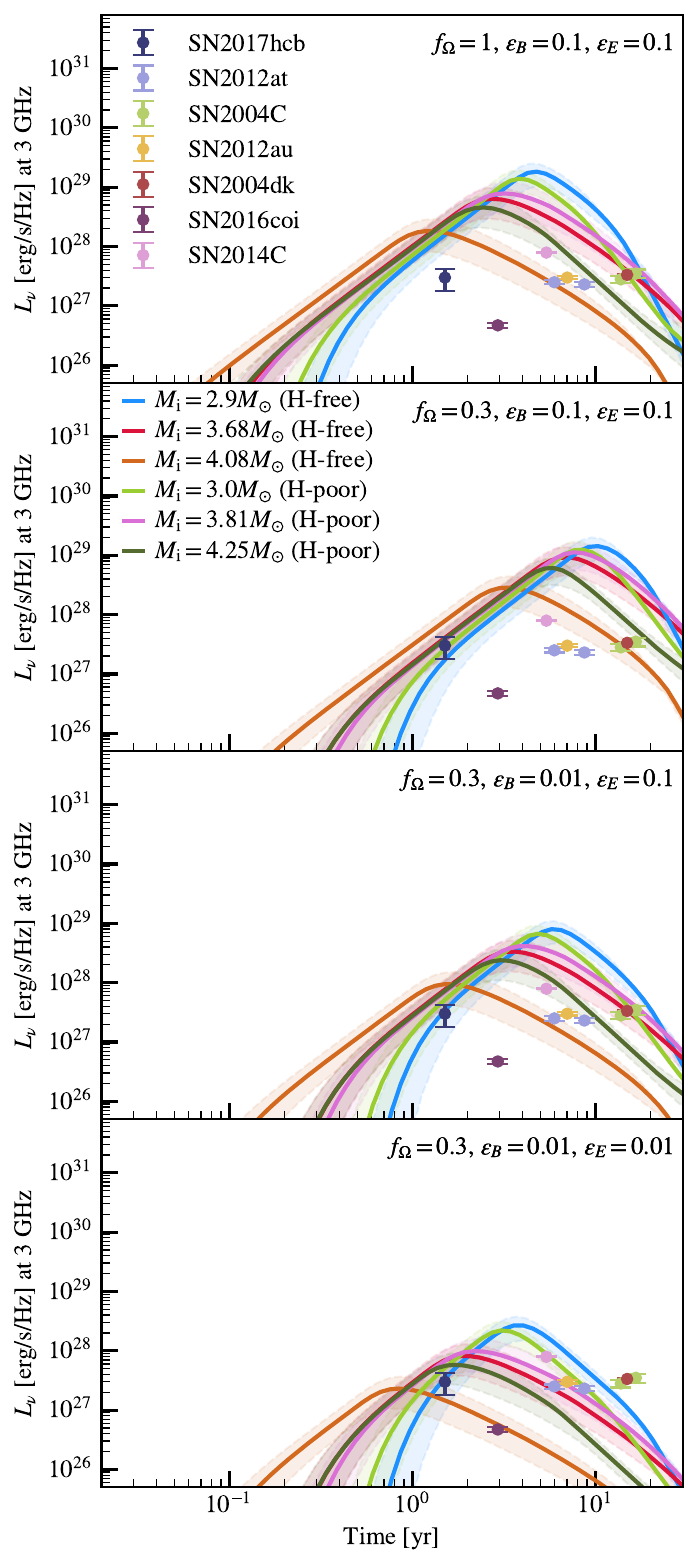}
    \caption{Light curves of radio emission at 3 GHz for the models listed in the legend of the second panel from the top. Each is calculated using the values of $f_{\Omega}$, $\epsilon_B$, and $\epsilon_E$ listed in the upper right of each panel. The solid lines represent the light curves for models assuming a kinetic energy of $E_{\rm ej} = 10^{51}$ erg. The shaded regions represent the range of emitted flux between $E_{\rm ej} = 5\times10^{50}$ erg and $E_{\rm ej} = 1.5\times10^{51}$ erg, with lower luminosities for smaller $E_{\rm ej}$. Scatter points represent observed late-time radio emission from a sample of events listed in the legend of the top panel, taken from \citep{Stroh21}. }
    \label{fig:allmodels_latetime}
\end{figure}

Figure \ref{fig:allmodels_latetime} shows the light curves at 3 GHz for our models, with the shaded region representing the spread in brightness as $E_{\rm ej}$ is varied from $5\times10^{50}$ erg to $1.5\times10^{51}$ erg. In the top panel of Figure \ref{fig:allmodels_latetime}, we show the resulting light curves if we assume a spherical CSM with $f_{\Omega}=1$, with $\epsilon_B=\epsilon_E=0.1$. The radio light curves rise to peak at 1--10 years after core collapse, with peak luminosities of $L_{\nu} \sim 10^{28}$--$10^{29}~\mathrm{erg}\, \mathrm{s}^{-1}\, \mathrm{Hz}^{-1}$. 

The second panel of Figure \ref{fig:allmodels_latetime} shows the results for an asymmetric CSM of $f_{\Omega}=0.3$ and with $\epsilon_B=\epsilon_E=0.1$. Compared to the spherical CSM case, the light curves for an aspherical CSM geometry peak later, since decreasing $f_{\Omega}$ produces a higher upstream CSM density (i.e. greater absorption) and lower $v_{\rm sh}$ at a given time $t$. The magnitude of the peak radio emission is only slightly brighter, as the increased magnetic field strength ($\propto \rho_{\rm CSM}^{1/2}$; see Equation \ref{eq:magnetic_field}) and synchrotron cooling for smaller values of $f_{\Omega}$ are mostly balanced by corresponding decreases in the power from ejecta-CSM interaction. 
Smaller values of $f_{\Omega}$ also correspond to increased FFA for viewing angles passing through the CSM as assumed here, along with increased SSA from the greater magnetic field strengths; these effects delay the peak time.

The third panel of Figure \ref{fig:allmodels_latetime} shows the results from assuming a lower magnetic field amplification efficiency $\epsilon_B=0.01$, while holding $f_{\Omega}=0.3$ and $\epsilon_E=0.1$. These light curves tend to be less luminous and peak earlier than in the second panel of Figure \ref{fig:allmodels_latetime} due to the reduction in synchrotron cooling and SSA for a lower $\epsilon_B$.
By lowering the electron acceleration efficiency $\epsilon_E$ as well to $\epsilon_E=0.01$, we produce light curves that are even dimmer and peak earlier, as seen in the bottom panel of Figure \ref{fig:allmodels_latetime}. These effects both result from the reduced number of relativistic electrons for lower $\epsilon_E$.

For our models of CSM originating from binary interaction, an asymmetric CSM structure is expected, with mass lost through the L2 point as a circumbinary torus \citep{pejcha2016,MacLeod2018}. The emission that would be observed before our predicted peak is affected by absorption processes that are sensitive to both the viewing angle and the ionization in the ejecta and CSM. As these dependencies are not incorporated into our forward modeling, we caution that there are large uncertainties in the pre-peak light curve behavior at $\lesssim1$ yr for our asymmetric models with $f_{\Omega} < 1$. We discuss this in greater detail in Section \ref{sec:viewingangles}.

In summary, our forward-modeling of CSM produced via non-conservative stable mass transfer predicts radio emission that peaks at 1--10 years from explosion, with luminosities at 3 GHz of $\approx10^{27}$--$10^{29}$ erg s$^{-1}$ Hz$^{-1}$ for values of ($\epsilon_E,\epsilon_B$) typically assumed in the literature.  

\subsection{Comparison to late-time radio-luminous SNe}
\label{sec:latetimeradio}
\cite{Stroh21} present a sample obtained by cross-matching of SNe with radio samples from the Very Large Array Sky Survey (VLASS), in which luminous radio emission of $L_{\nu} \sim 10^{26}$--$10^{29}~\mathrm{erg}\, \mathrm{s}^{-1}\, \mathrm{Hz}^{-1}$ is detected at 2--4 GHz from years to decades after core-collapse (see also \citealt{Rose24} for similar work using ASKAP). The majority of the sample is characterized as H-poor from early optical spectra, indicating that their progenitors have lost some or all of their H envelope at the time of core collapse. \cite{Stroh21} suggest that the bright late-time radio emission may be interpreted as interaction between the SN shock and dense CSM formed via large mass loss rates (e.g., $\dot{M}\gtrsim 10^{-4} M_{\odot}\, \rm{yr}^{-1}$ for their assumed wind velocity of $10^3~\mathrm{km}\ \mathrm{s}^{-1}$).

\begin{figure}
    \centering
    \includegraphics[width=\columnwidth]{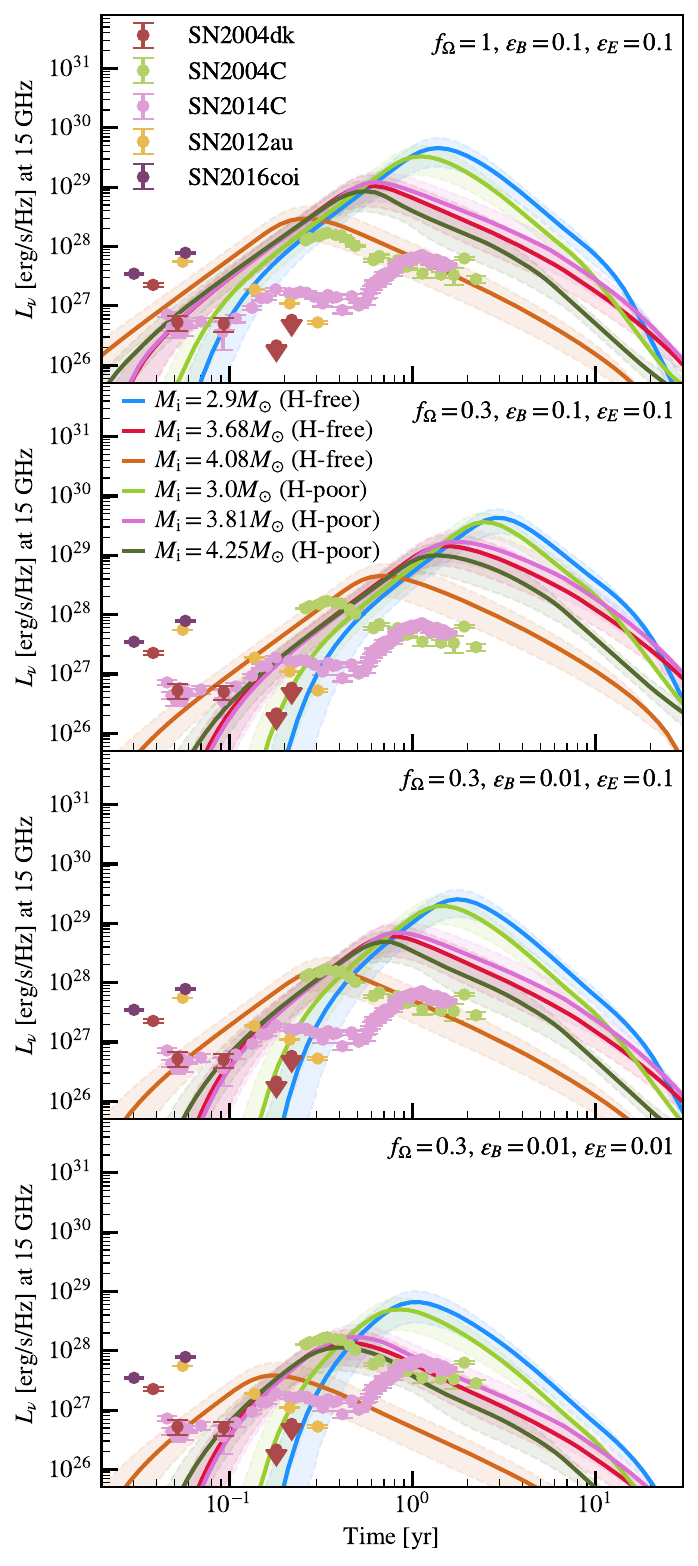}
    \caption{Light curves of radio emission at 15 GHz for the same models as in Figure \ref{fig:allmodels_latetime}. Scatter points show observed radio emission at $\approx15$ GHz for the events listed. }
    \label{fig:allmodels_earlytime}
\end{figure}

\begin{figure*}
    \centering
    \includegraphics[width=\linewidth]{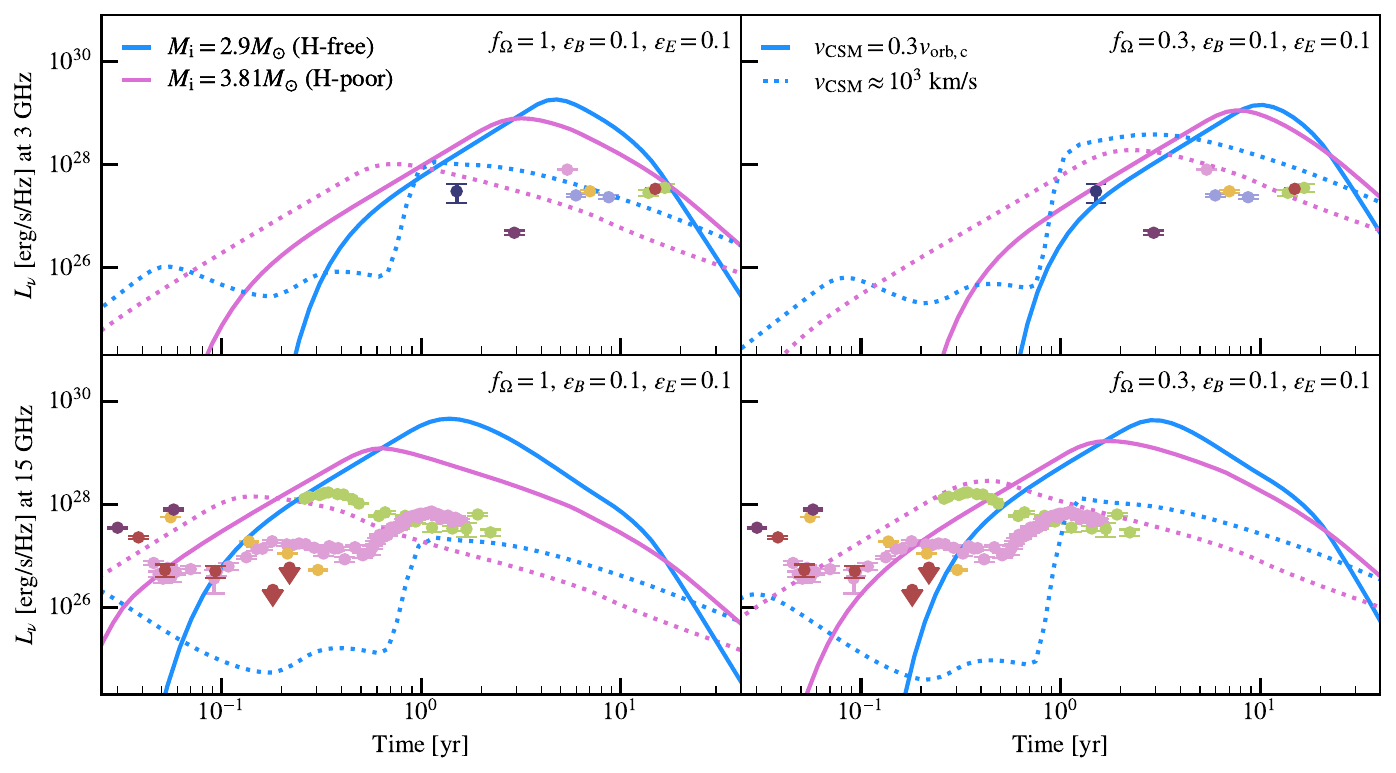}
    \caption{\textit{Top row}: Light curves of radio emission at 3 GHz for the models listed in the legend. Each is calculated with $E_{\rm ej}=10^{51}$ erg and the parameters listed in the upper right of each panel.
    The solid lines show the light curve for a CSM velocity of $0.3\, v_{\rm orb, c}$, as in Figure \ref{fig:allmodels_latetime}. The dotted lines show the light curve for a fast CSM of $10^3~\mathrm{km}\ \mathrm{s}^{-1}$, assuming the CSM has been accelerated by some mechanism upon ejection from the binary system. The scatter points represent the same events as listed in the legend of Figure \ref{fig:allmodels_latetime}. \textit{Bottom row}: Light curves of radio emission at 15 GHz for the same models as in the top row. The scatter points represent the same events as listed in the top right legend of Figure \ref{fig:allmodels_earlytime}.
    }
    \label{fig:compare_v_csm}
\end{figure*}

We compare our predicted luminosity to the observed late-time radio emission from a subset of H-poor events in the \cite{Stroh21} sample, shown as scatter points in Figure \ref{fig:allmodels_latetime}. The sample serves as a benchmark for the typical range of radio luminosities that has been detected at late times subsequent to H-poor SNe. Four events noted as H-poor in the sample are associated with gamma-ray bursts or classified as broad-lined Ic SNe, so these are likely engine-powered events with much larger inferred $E_{\rm ej}$; we exclude these events from the observed SNe shown in Figures \ref{fig:allmodels_latetime}--\ref{fig:viewing_angles}. We note that the outer layers of our stripped stars remain He-rich at core collapse, so they likely give rise to Type Ib rather than Type Ic SNe; nevertheless, for completeness we still show events classified as Type Ic among the scatter points, as they are similar in luminosity to the Type Ib samples.  Even for our most pessimistic assumptions of $\epsilon_B$ and $\epsilon_E$, the luminosities achieved by our models are as bright as the observed late-time radio emission during the time period of a few--10 years after core collapse.

\cite{Stroh21} propose several scenarios to power the luminous late-time radio emission in their sample, noting that interaction with dense, detached CSM shells is a likely candidate to explain the observed radio data for at least some of the samples (e.g., SN 2004dk, 2012au, 2014C). 
Of the CSM models produced by binary interaction in this work,  we predict that detached CSM can generally be produced by the $M_i = 2.9\, M_{\odot}$ (H-free) and $M_i = 3.0\, M_{\odot}$ (H-poor) models, as these low-mass stripped star progenitors detach from their Roche lobes in the last few years before core collapse (Figure \ref{fig:density}). However, the modeled CSM velocities of $\lesssim 100$ km s$^{-1}$ result in a shell which is separated by only $\lesssim$ a few $\times 10^{15}$ cm, with a monotonic wind-like profile exterior to this radius. Thus, the light curves in Figure \ref{fig:allmodels_latetime} still rise over a few years to peak.
We note that for events where the radio luminosity remains nearly constant between multiple epochs, \cite{Stroh21} also discuss the possibility of the emission originating from GRB jet propagation through a wind-like density profile, which can manifest as a flat peak \citep{granot2018}; however, \cite{Stroh21} also deem relativistic jets to be unlikely for the events we have included here.

Overall, our models represent stripped stars with masses in the expected range for typical Type Ib SN progenitors, and the continuous structure of the CSM for our higher-mass models originates self-consistently from placing these stars in short-period binaries and tracking their non-conservative mass transfer. Our models therefore demonstrate that dense CSM, either wind-like or detached depending on the progenitor system, is a likely outcome for the stripped star progenitors of H-poor SNe that exist in close binaries. In turn, the subsequent interaction of the SN ejecta with the dense CSM is able to power highly luminous late-time radio emission that is observable with surveys like VLASS.

\subsection{Comparison to existing early-time radio data}
\label{sec:early_data}

A subset of the VLASS-detected events shown in Figure \ref{fig:allmodels_latetime} have also been studied in the radio from weeks to years after the SN. In Figure \ref{fig:allmodels_earlytime}, we show some of the early-time data at 15 GHz for SN 2004dk \citep{wellons2012} and SN 2004C \citep{demarchi2022}, as well as at 15.7 GHz for SN 2014C \citep{anderson2017}, at 14.75 GHz for SN 2016coi \citep{terreran2019}, and at 16 GHz for SN 2012au \citep{kamble2014}. We compare the observed radio emission to our model light curves at 15 GHz, again varying $E_{\rm ej}$ from $5\times10^{50}$ erg to $1.5\times10^{51}$ erg and exploring different assumptions for $f_{\Omega}$, $\epsilon_B$, and $\epsilon_E$. The model light curves at 15 GHz in Figure \ref{fig:allmodels_earlytime} systematically peak earlier than the light curves at 3 GHz in Figure \ref{fig:allmodels_latetime} and achieve similar peak luminosities of $L_{\nu}\approx 10^{27}$--$10^{30}~\mathrm{erg}\, \mathrm{s}^{-1}\, \mathrm{Hz}^{-1}$.

The top panel of Figure \ref{fig:allmodels_earlytime} shows the light curves for $f_{\Omega}=1$ and $\epsilon_B=\epsilon_E=0.1$. With the exception of the $M_i=4.08\, M_{\odot}$ (H-free) model, the model light curves peak much later than the observed radio emission for these events. Even then, the $M_i=4.08\, M_{\odot}$ (H-free) model bears similarity only to the early radio data of SN 2004C, which also peaks on a timescale of a few months; nevertheless, SN 2004C is much brighter in the late-time VLASS epoch than this progenitor model predicts (Figure \ref{fig:allmodels_latetime}). SN 2004dk and SN 2012au each appear to show decline from peak emission at $\lesssim\, $weeks after the SN, and the early light curve of SN 2016coi seems to be rising over that duration as well---these timescales are far shorter than the rise to peak of any of our models. Finally, the early radio emission from SN 2014C exhibits multiple bumps, a significant departure from the shape of our single-peaked light curves. For these events with available early-time radio data, our $f_{\Omega}=1$, spherically symmetric, models are in tension with the observed radio emission.

The light curves from asymmetric, $f_{\Omega}=0.3$ models with different assumptions of $\epsilon_B$ and $\epsilon_E$ (as shown in the bottom three panels of Figure \ref{fig:allmodels_earlytime}) similarly do not replicate the decline from an early peak apparent for SN 2004dk and SN 2012au, the rise to an early peak for SN 2016coi, or the multiple peaks of SN 2014C. While the peak timescale and luminosity of SN 2004C is not dissimilar from some of the model light curves with $\epsilon_B =0.01$, comparison with Figure \ref{fig:allmodels_latetime} shows that no model self-consistently explains both the early- and late-time emission from SN 2004C.

However, the smooth rise to and decline from a bright peak exhibited by the model light curves are in all likelihood oversimplified compared to reality. 
As mentioned in Section \ref{sec:synchrotron} and earlier in Section \ref{sec:results}, in the case of asymmetric CSM with $f_\Omega<1$, significant uncertainty underlies emission from polar regions that are not subtended by the dense circumbinary CSM.  Our treatment of absorption does not account for differences when the system is viewed from polar angles, which can be important at early times. The polar regions may also be filled by more diffuse CSM, e.g. a stellar wind from the progenitor and possibly the companion. Interaction between the SN ejecta and this low-density material can power an additional early peak in the radio light curve. We explore this prospect in Section \ref{sec:alternatives} and elaborate on the uncertainties in Section \ref{sec:uncertainties}. 

\subsection{Can radio emission from the stable mass transfer scenario explain both early and late observations?}
\label{sec:alternatives}

Overall, our forward modeling of CSM from non-conservative stable mass transfer naturally reproduces observations of bright late-time radio emission, as demonstrated in Figure \ref{fig:allmodels_latetime}. However, under the assumptions used thus far, the models struggle to self-consistently explain early-time radio data from a subset of events, which suggest a non-monotonic rise in the radio light curves before the late-time emission detected in VLASS.

To reproduce the characteristics of the early-time data likely requires incorporating further complexities into our models of the CSM density distribution. For example, the CSM could be moving faster than our expectation of typical ejection speeds $v_{\rm CSM}\sim 100$ km s$^{-1}$. A higher $v_{\rm CSM}$ leads to the following effects, which may occur in tandem: firstly, the radio emission can emerge earlier due to the reduced density ($\propto v_{\rm CSM}^{-1}$); secondly, detached CSM (as seen in our $M_i = 2.9\, M_{\odot}$ (H-free) and $M_i = 3.0\, M_{\odot}$ (H-poor) models) is propelled to larger distances from the progenitor, creating an extended, low-density wind bubble that potentially appears as early emission. 

A further possibility is that the CSM can be described by a two-component structure composed of the torus of aspherical dense CSM and a separate lower-density CSM component, such as the progenitor's stellar wind. In this scenario, emission $\lesssim\ $weeks after the SN may be powered by interaction with the stellar wind. This framework can apply to the wind-like CSM profiles as well, without necessarily invoking the highly detached CSM distribution that has been proposed in prior literature \citep[e.g.,][]{Stroh21}.

In either case, different viewing angles of the progenitor system may influence the appearance of the early-time light curve, but such effects are difficult to capture fully without a more sophisticated framework that includes viewing-angle dependencies in the calculations for shock propagation and radio absorption. Nevertheless, we may still examine the viability of these two alternatives within our current framework. We consider a faster CSM velocity in Section \ref{sec:fastcsm} and another viewing angle for asymmetric CSM in Section \ref{sec:viewingangles}.

\subsubsection{Varying the CSM ejection velocity} 
\label{sec:fastcsm}

Dense, detached CSM located further from the progenitor may produce light curves characterized by early peaks and re-brightening at late times.
Faster CSM velocities than $v_{\rm CSM} = 0.3\, v_{\mathrm{orb},c}\; \lesssim 100$ km s$^{-1}$, as assumed thus far, can shift the detached shell of the $M_{\rm i}=2.9M_{\odot}$ (H-free) or $M_{\rm i}=3.0M_{\odot}$ (H-poor) model's CSM profile to larger radii and create a low-density cavity, as suggested in \cite{Stroh21} for some of the VLASS SNe. 
A larger CSM velocity may be pertinent in the case of a NS companion, where the high mass transfer rates we see in our binary models will lead to super-Eddington accretion. As explored in \cite{Tsuna2024}, interaction between the ensuing fast disk wind and the circumbinary torus of dense CSM can accelerate the CSM to much larger velocities of up to $\sim 10^{3}~\mathrm{km}\ \mathrm{s}^{-1}$, so that the CSM at the time of the SN is more extended by a factor of $\sim10$.

In Figure \ref{fig:compare_v_csm}, we explore assumptions for a faster CSM velocity in two of our models, the $M_i=2.9~M_\odot$ (H-free) model, which exhibits a detached shell of CSM, and the $M_i=3.8~M_\odot$ (H-poor) model, which we select as representative of a wind-like CSM profile. The 3 GHz light curves are shown for $f_{\Omega}=1$ on the top left and for $f_{\Omega}=0.3$ on the top right, both with $\epsilon_B=\epsilon_E=0.1$. The solid curves represent models with $v_{\rm CSM}=0.3\, v_{\rm orb,c}$, whereas the dotted light curves are from CSM density profiles calculated with $v_{\rm CSM}\approx 10^3~\mathrm{km}\ \mathrm{s}^{-1}$. For both types of CSM profiles, the light curves assuming faster CSM are dimmer than the slower CSM models because the density at a given radius is vastly decreased as the CSM is spread out across larger radii.

In the case of $v_{\rm CSM}\approx 10^3~\mathrm{km}\ \mathrm{s}^{-1}$, the $M_i=3.81\, M_{\odot}$ (H-poor) model light curve peaks earlier, since the lower CSM densities cause vastly decreased absorption. For both values of $f_\Omega$, the light curves of the $M_i = 2.9\, M_{\odot}$ (H-free) model rise sharply at $\sim 1$ yr after the SN to peak brightnesses of $L_{\nu}\sim 10^{28}~\mathrm{erg}\, \mathrm{s}^{-1}\, \mathrm{Hz}^{-1}$. The timing of the rise is set by when the SN ejecta reaches the detached dense CSM shell. This differs from the model with lower $v_{\rm CSM}$, in which the rise is set by the decrease of the absorption optical depths over time.

In addition, the $M_i = 2.9\, M_{\odot}$ (H-free) model with $v_{\rm CSM}\approx 10^3~\mathrm{km}\ \mathrm{s}^{-1}$ shows two earlier peaks of $L_{\nu}\lesssim10^{26}~\mathrm{erg}\, \mathrm{s}^{-1}\, \mathrm{Hz}^{-1}$, on timescales of $\lesssim $ weeks and $\sim$ months respectively. This early-time emission is produced by shock interaction with the low-density stellar wind interior to the dense CSM shell.
We note that the second peak at $\sim 0.3$ yr corresponds to a slight uptick in the wind mass loss rate of the stripped progenitor. While the magnitude of this bump may vary for different wind mass loss prescriptions, it is an interesting feature that could appear for detached CSM profiles.

In the bottom row of Figure \ref{fig:compare_v_csm}, we compare the predicted radio light curves at 15 GHz for the same model assumptions to the early-time data for the events shown in Figure \ref{fig:allmodels_earlytime}. For the wind-like CSM of the $M_i=3.81\, M_{\odot}$ (H-poor) model, the faster CSM velocity causes the peak timescale and brightness of the light curve to more closely resemble the observed radio data from SN 2004C at early times. In particular, when using $v_{\rm CSM}\approx 10^3~\mathrm{km}\ \mathrm{s}^{-1}$, the light curve assuming $f_{\Omega}=0.3$ is consistent with the observed emission from SN 2004C at both early and late times. The rise to peak also proceeds with a similar timescale and slope as the early radio data for SN 2016coi, though the model light curve is not as bright.

For the $M_i = 2.9\, M_{\odot}$ (H-free) model at 15 GHz, the timescales of the early peaks when assuming $v_{\rm CSM}\sim 10^3~\mathrm{km}\ \mathrm{s}^{-1}$ are also similar to those of the observed peaks in the early radio data for SN 2004dk, SN 2012au, and SN 2014C. However, the model light curve is systematically dimmer than the observed early emission from these events. 
Given the uncertainties in the wind mass-loss rates of these stars near core-collapse, the discrepancy may be reconciled with a larger pre-SN $\dot{M}_{\rm w}$ by a factor of $\gtrsim 10$ than adopted in Section \ref{sec:massloss} (e.g. \citealt{Sander20,Moriya22}; see also Section \ref{sec:uncertainties}).

Our forward modeling demonstrates that by assuming $v_{\rm CSM}\sim 10^3~\mathrm{km}\ \mathrm{s}^{-1}$, interaction with a density profile akin to the $M_i = 2.9\, M_{\odot}$ (H-free) model produces light curves that may be relevant to similar events showing radio re-brightening after dimmer emission during the first months--years after the SN. Due to the lower densities in the faster-moving CSM, the bright late-time emission is reproduced with larger, but still reasonable, values of $\epsilon_B$ and $\epsilon_E$. In addition, comparison to the early-time peaks in the radio emission favors a denser stellar wind from the stripped star progenitor, which may well be present from the late stages of stellar evolution that is probed during such early phases of the SN \citep{Gilkis2025,Maeda22}.

\begin{figure*}
    \centering
    \includegraphics[width=\linewidth]{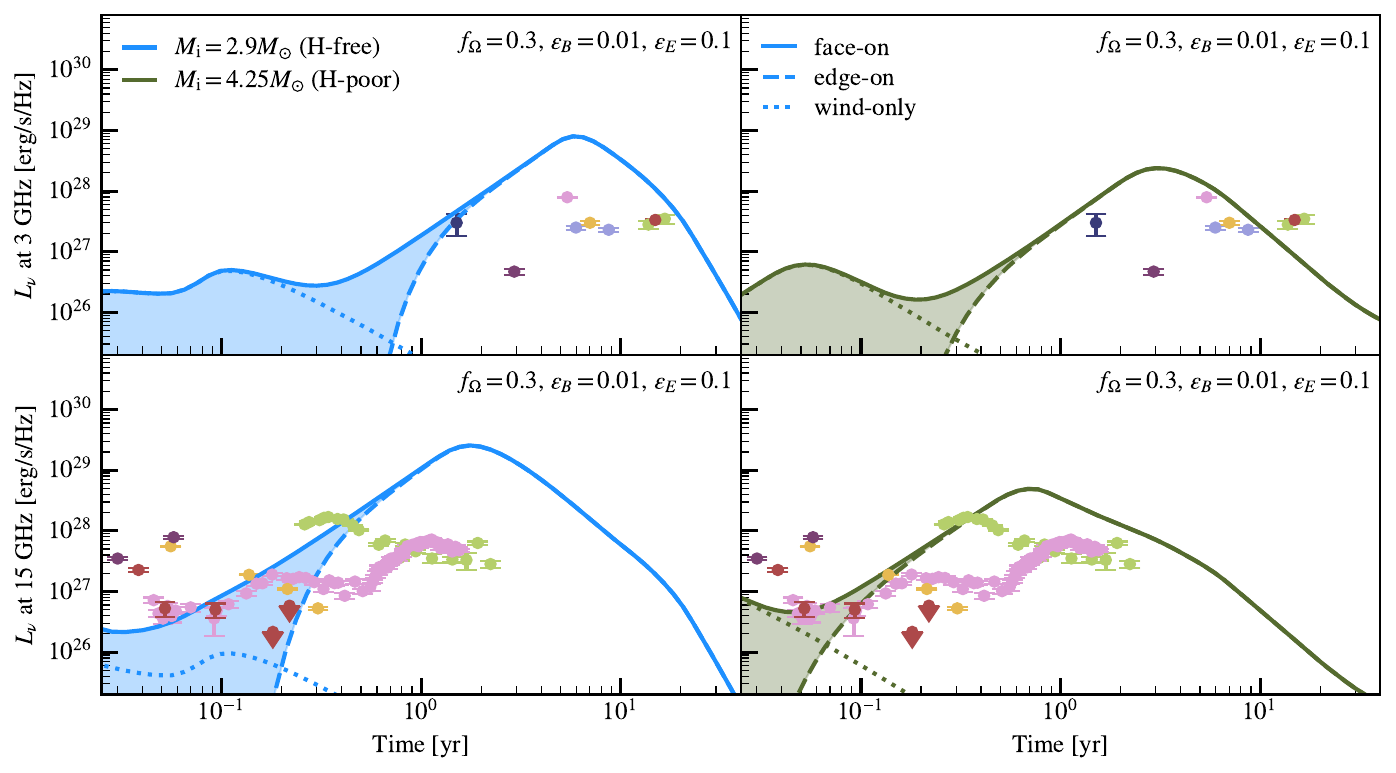}
    \caption{\textit{Top row}: Light curves of radio emission at 3 GHz for the models listed in the legend. Each is calculated with $E_{\rm ej}=10^{51}$ erg and the parameters listed in the upper right of each panel.
    The solid lines show an estimate of the light curve for a face-on viewing angle, which is the sum of contributions from the low-density stellar wind in the polar regions and from the dense CSM in the torus. The early peaks appear due to the low-density stellar wind (dotted line), and the second bright peak is set by SSA for the CSM.  The dashed lines show the light curve from an edge-on viewing angle along the direction of the CSM, as in Figure \ref{fig:allmodels_latetime}, which represents the maximum absorption due to both FFA and SSA. The shaded region encompasses the range of emission between the face-on and edge-on viewing angles. The scatter points represent the same events as listed in the legend of Figure \ref{fig:allmodels_latetime}. \textit{Bottom row}: Light curves of radio emission at 15 GHz for the same models as in the top row. The scatter points represent the same events as listed in the top right legend of Figure \ref{fig:allmodels_earlytime}.
    }
    \label{fig:viewing_angles}
\end{figure*}

\subsubsection{Another viewing angle of the asymmetric CSM}
\label{sec:viewingangles}

In the case of $f_\Omega<1$, we consider the CSM to be asymmetric and distributed as a circumbinary torus. Our framework for synchrotron emission and absorption has thus far effectively assumed an edge-on viewing angle for this torus, in which the radio light curve is observed along the direction of the CSM (see discussion at the end of Section \ref{sec:synchrotron}). Here we explore how the predicted radio light curve at early times may differ for an observer viewing the torus face-on.

For a face-on viewing angle, the interaction between the SN shock and the stellar wind occupying the region not subtended by the torus will contribute to the synchrotron emission at early times. We calculate the radio emission from this component for the face-on case. In addition, we assume that when observing the system face-on, the interaction with the dense CSM torus will be effectively attenuated only by SSA with negligible effects from FFA, since the line-of-sight no longer passes through the dense CSM. The final light curve is calculated as the sum of these two emission components.

Figure \ref{fig:viewing_angles} illustrates how the viewing angle of the CSM influences the early-time light curve. Here, we show the $M_{\rm i}=2.9\, M_{\odot}$ (H-free) and $M_{\rm i}=4.25\, M_{\odot}$ (H-poor) models for the parameters listed in the top right of each panel. Our modeled radio light curves using these parameter choices are shown in Figure \ref{fig:allmodels_latetime} to more closely resemble the observed late-time radio emission from the \citet{Stroh21} sample at 3 GHz. The dashed lines in each panel represent the edge-on case and are equivalent to the third panel in each of Figures \ref{fig:allmodels_latetime} and \ref{fig:allmodels_earlytime}. The solid lines show our models for a face-on viewing angle, which consists of contributions from the lower-density stellar wind at early times (also shown as dotted lines) and from the dense CSM torus at late times. Since for the face-on case we neglect FFA, the light curve from interaction with the dense torus rises at earlier times than for the edge-on case. However, since the peak of the late-time emission is governed by SSA, the ability of our models to reproduce the observed late-time emission remains unaltered by viewing angle effects.

For a face-on viewing angle, an early peak powered by interaction with the stellar wind is visible on timescales of weeks to months after the explosion for the 3 GHz light curves (top row of Figure \ref{fig:viewing_angles}). At 15 GHz, the interaction with the stellar wind produces earlier and dimmer peaks. Similar to the fast CSM scenario explored in Section \ref{sec:fastcsm}, the properties of the early peak depend on the strength of the stellar wind and our chosen wind mass loss prescription. Comparison to the early-time data would again favor a denser stellar wind from the stripped star progenitor in order to produce brighter emission at 15 GHz, which may be plausible shortly before core collapse \citep{Gilkis2025}.

Our exercise in predicting the emission and absorption for a face-on viewing angle is highly idealized, and more work is needed to construct a framework that can capture the dependence of the light curve on viewing angle. The early-time light curve for viewing angles between the two extremes we have examined here may take on intermediate values within the shaded region in each panel of Figure \ref{fig:viewing_angles}, which represents the range of radio emission between the face-on and edge-on light curves. Our exploration of the face-on case demonstrates that different viewing angles can reveal contributions from the progenitor's stellar wind manifesting as early peaks, which are qualitatively similar to the observed early-time radio data. At late times, interaction with the dense CSM still produces luminous radio light curves comparable to the observed late-time radio emission.

\section{Discussion}
\label{sec:discussion}
\subsection{Early signatures of interaction with dense CSM}
Early-time optical evidence of interaction does not appear for the majority of the sampled VLASS events \citep{Stroh21}. Since the CSM in our models is denser than typical Wolf-Rayet winds, in this section we assess whether the dense CSM in our models would lead to observable interaction signatures in early-time optical data.

For the detached, fast CSM assumption explored in Section \ref{sec:fastcsm}, an absence of early interaction signatures is expected, as the dense CSM lies at larger distances. Alternate viewing angles of asymmetric CSM, as in Section \ref{sec:viewingangles}, could also be consistent with unseen interaction signatures: a lack of line emission at early times is natural for a torus-like CSM ($f_{\Omega} < 1$), since the rapidly expanding SN ejecta in the less dense polar region can mask the interaction signatures occurring below its photosphere \citep[e.g. Figure 2 of][]{smith2017}. This obscuration is effective during the  first months of the SN in its photospheric phase, where the ejecta's optical depth $\tau_{\rm ej}\approx 3\kappa M_{\rm ej}/4\pi(v_{\rm ej}t)^2\sim 7\, (t/50\ {\rm day})^{-2}(M_{\rm ej}/2~M_\odot)^2(E_{\rm ej}/10^{51}\ {\rm erg})^{-1}$ is at least a few. Here $\kappa\approx 0.07\ {\rm cm^2\ g^{-1}}$ is the opacity to optical emission \citep{Taddia18}, and $v_{\rm ej}=\sqrt{2E_{\rm ej}/M_{\rm ej}}$ is the ejecta's bulk velocity.

Signatures of CSM interaction may also appear in the early-time optical light curve. In our spherically symmetric models in Figure \ref{fig:density}, the CSM densities at distances of $\sim 10^{14}$--$10^{15}$ cm are at least an order of magnitude lower than those of Type Ibn SNe that show interaction signatures in early-time optical data \citep{Maeda22}, indicating that the absence of such signatures is plausible for our models in that region. Meanwhile, the density profiles of the $M_i = 2.9\, M_{\odot}$ (H-free) and $M_i = 3.0\, M_{\odot}$ (H-poor) models do overlap at $\gtrsim 3\times10^{15}$~cm with the density range derived in \cite{Maeda22}. 
For asymmetric CSM ($f_{\Omega}< 1$), our model densities are a factor of $f_\Omega^{-1}$ higher and can approach what is inferred for SNe Ibn.

Nevertheless, we estimate that the luminosity powered by interaction with the CSM at distances of $\sim 10^{14}$--$10^{15}$ cm in our models is subdominant compared to the $^{56}$Ni-powered emission in the first days--weeks of the SN. In this region, the CSM density can be approximated 
as a $\rho \propto r^{-2}$ wind, with $\rho_{\rm CSM} \sim \rho_0\, \left(r/10^{15} \mathrm{cm}\right)^{-2} (f_\Omega/0.3)^{-1}$, with $\rho_0\approx 10^{-16}\, \mathrm{g}\, \mathrm{cm}^{-3}$ for all but the $M_i = 2.9\, M_{\odot}$ (H-free) and $M_i = 3.0\, M_{\odot}$ (H-poor) models. The luminosity powered by interaction between this CSM and the forward shock is given by 
\begin{eqnarray}
    L_{\rm CSM} &=& 2\pi r^2f_\Omega \rho v_{\rm sh}^3 \\
                &\approx& 6\times10^{41}\, \mathrm{erg/s}\, \left(\frac{\rho_0}{10^{-16}\, \mathrm{g}\, \mathrm{cm}^{-3}}\right) \left(\frac{v_{\rm sh}}{10^4\, \mathrm{km}\, \mathrm{s}^{-1}}\right)^3, \nonumber 
\end{eqnarray}
ultimately independent of $f_\Omega$.

At a given time $t$, the actual contribution to the light curve from interaction is given by $L_{\rm int} \sim L_{\rm CSM}\, {\rm min}(t/t_{\rm cool},1)$, where $t_{\rm cool}$ represents the timescale to cool the shock-heated gas by converting its internal energy into photons. Given the emissivity $\epsilon_{\rm cool}$ from Equation \ref{eq:emissivity},
\begin{eqnarray}
     t_{\rm cool} &=& \frac{1.5 (n_e+n_i)k_B T}{\epsilon_{\rm cool}} \\\
     \approx &10^{3}& \mathrm{d} \left(\frac{v_{\rm sh}}{10^4\, \mathrm{km}\, \mathrm{s}^{-1}}\right)^3\! \left(\!\frac{t}{10\, \mathrm{d}}\!\right)^2 \!\left(\!\frac{\rho_0}{10^{-16}\, \mathrm{g}\, \mathrm{cm}^{-3}}\!\right)^{-1} \!\left(\!\frac{f_{\Omega}}{0.3}\!\right) . \nonumber 
\end{eqnarray}

Thus, on a timescale of $t\gtrsim 10\, \mathrm{d}$, the potential contribution to the optical light curve from interaction between the SN shock and the CSM in our wind-like models is $L_{\rm int} \lesssim 10^{40}\, \mathrm{erg}\,\mathrm{s}^{-1}$, using $f_{\Omega}=0.3$ and $\rho_0=10^{-16}\, \mathrm{g}\, \mathrm{cm}^{-3}$. This is negligible compared to the $^{56}$Ni-powered emission of $\sim 10^{42}\, \mathrm{erg}\, \mathrm{s}^{-1}$ \citep{Taddia18}. 
As the $M_i = 2.9\, M_{\odot}$ (H-free) and $M_i = 3.0\, M_{\odot}$ (H-poor) models have $\rho_0\sim 2 \times10^{-16}\, f_{\Omega}^{-1}\, \mathrm{g}\, \mathrm{cm}^{-3} $ at $\gtrsim 2\times10^{15}$ cm, the emission from interaction at $\sim1$ week can potentially reach $\lesssim 10^{41}\, \mathrm{erg}\,\mathrm{s}^{-1}$ and add a $\lesssim 10\%$ contribution to the peak brightness.
The magnitude of the contribution will further decrease if not all photons are efficiently converted to the optical, as expected for the densities predicted here \citep[e.g.,][]{Margalit22}. 

\subsection{Other signatures of CSM interaction} 
While early signatures of CSM interaction are generally absent in the VLASS SNe, late-time line-emission signatures have been seen for some events, supporting the CSM scenario. For events such as SN 2014C \citep{Margutti17} and SN 2004dk \citep{pooley2019}, H$\alpha$ lines likely due to ejecta-CSM interaction have been identified at late times. Other examples include the radio transient VT J121001+495647 \citep{Dong21}, which was also posited to originate from collision of SN ejecta with dense, asymmetric CSM ejected via binary mass loss during the inspiral leading to merger. 

Our stripped star models are nearly free of hydrogen, so reproducing H$\alpha$ emission is not straightforward. Even in our H-poor models that contain $\sim 0.01\,M_{\odot}$ of hydrogen, the H-rich material is stripped early in the binary evolution so that it lies at $\gtrsim 10^{18}$ cm at core-collapse. For interaction with this H-rich material to occur by $10$--$20$ yr would require sustained shock velocities of $\sim 0.1c$, at least a few times faster than the typical shock velocity at $\gtrsim 1$ yr in our models. 
For more representative shock velocities, the SN ejecta would instead be interacting with He-rich CSM (with a mass fraction of $Y_{\rm He} \gtrsim 0.8$) on year to decade timescales after the SN, so our models would predict spectra during that period to be devoid of H signatures while possessing He features. SNe with evidence of H$\alpha$ during that period require an explanation that can eject H-rich material closer to core collapse.

The presence of hydrogen may originate from a main-sequence (MS) companion, and studying the evolution of similar short-period binaries with MS accretors will be important in the future. For a MS star, non-conservative mass transfer may ensue if the mass accretion can inflate the companion's stellar envelope enough for the companion to overflow its Roche lobe. Models of rapid accretion at mass transfer rates similar to those of our work show the star to inflate on short timescales of $10^2$--$10^3$ yr and remain inflated for $\approx10^5$ yr
\citep[][]{Lau24}. 
This process could also induce convective mixing,  
which could result in hydrogen-rich CSM if the convection is able to efficiently mix the H-rich envelope of the MS star with the He-rich accreted material.
Simulating the accretion of He-rich material onto a MS star and the subsequent evolution, similar to the methods of \cite{Lau24}, would be a worthwhile future avenue to explore this possibility.

\subsection{Other binary interaction scenarios}
\label{sec:otherscenarios}

In addition to the scenario of stable mass transfer from a low-mass stripped star that we study in this work, other forms of binary interaction can also create dense CSM. Of particular interest are systems that can create detached CSM located at large radii, a morphology which our results in Section \ref{sec:fastcsm} and the inference of earlier work \citep[e.g.,][]{Stroh21} indicate is well-suited to explain the observed radio emission from the sampled events. We discuss a subset of these possibilities below.

Stars with more massive H-rich layers than our modeled progenitors, such as supergiants in wide binaries that undergo Case C mass transfer \citep{ercolino2024,matsuoka2024}, can lose mass on timescales compatible with those suggested by observed late-time radio emission. While this process can lead to H-rich CSM at large distances, 
stable case C Roche-lobe overflow alone is not likely to fully remove the supergiant envelope, such that a H-poor SNe ensues upon core-collapse. Low-metallicity stripped stars can also retain up to a few$\,\times\, 0.1\, M_{\odot}$ of H-rich envelope until as late as core He depletion \citep{Gotberg17,laplace2020}; thus, their binary interactions on core carbon burning timescales may also produce H-rich CSM, while resulting in a H-poor SN. However, the effect of metallicity is less relevant to events hosted in near-solar regions \citep[e.g. SN 2014C and SN 2004dk,][]{Ganss22}. 

In some cases, Case C mass transfer in a wide binary system can be unstable and lead to common envelope ejection of the supergiant's envelope. Common envelope would naturally lead to detached H-rich CSM and H-poor SNe, provided that the bulk of the envelope mass $M_{\rm CE}$ can be unbound with velocities of $v_{\rm CSM}\sim 10$--$100$ km s$^{-1}$ at $\sim 10^4$ yrs before core-collapse. This can form a shell with a density of $\sim 10^{-19}$--$10^{-18}\, \mathrm{g}\, \mathrm{cm}^{-3} \left(M_{\rm CE}/10 M_{\odot}\right) \left(v_{\rm CSM}/10\, \mathrm{km}\, \mathrm{s}^{-1}\right)^{-3}$, depending on the shell width. 
This possibility was suggested for SN 2014C, which displayed H$\alpha$ emission at a later phase \citep[e.g.,][]{Margutti17,Brethauer22,orlando2024}. 
The outcomes of binary systems undergoing this process and the morphology of the resultant CSM strongly depend on uncertain common envelope physics \citep[e.g.,][]{lau2022a,lau2022b,ropke2023,Lau25}, but this is a promising explanation that ought to be explored further in future work. 

Another possibility to produce detached CSM is stable mass transfer ending well before the donor's core collapse, if the transferred mass is able to leave the system. In binaries with much larger mass ratios ($q=M_{\rm c}/M_{\rm *}$) than considered in our work, the angular momentum evolution can lead to widening of the binary \citep{tauris2006,marchant2025}, thereby potentially terminating Roche-lobe overflow well before core-collapse. Such binaries may moreover be more common from a population standpoint  \citep[e.g.,][]{Zapartas17}, yet whether non-conservative mass transfer will manifest in such systems is also more uncertain. Since massive stars have shorter thermal timescales, mass transfer onto higher-mass main sequence accretors tends to be more conservative, and mass accretion at our typical rates of $\dot{M}\sim10^{-4}~M_\odot\ {\rm yr}^{-1}$ is not expected to significantly inflate the envelopes of main sequence stars with masses $\gtrsim 3~M_\odot$ \citep{Lau24}. However, the accreting massive stars may be spun up to large fractions of their critical rotation rates due to mass accretion, potentially leading to highly non-conservative mass transfer via rotationally-enhanced winds \citep[][and references therein]{rocha2024}. 

\subsection{Model uncertainties}
\label{sec:uncertainties}
In future work, several uncertainties in our models can be narrowed down. 
For instance, the velocity of the ejected CSM is not well constrained. The value $v_{\rm CSM}=0.3\, v_{\rm orb,c}$ we assume initially is based on a theoretical framework for the dynamics of the CSM ejection from the binary system. However, narrow lines associated with CSM that appear in interacting SN spectra typically exhibit line widths of $\sim 10^2$--$10^3~\mathrm{km}\ \mathrm{s}^{-1}$ \citep[][]{Pastorello08,hosseinzadeh2017,strotjohann2021}, which is an order of magnitude larger than the values of $v_{\rm CSM}$ in our models. Nevertheless, we note that the CSM probed by these lines can be accelerated by the SN radiation \citep{Tsuna2023}, and the speed of CSM leaving the system (before the SN) may actually be more consistent with the slower velocities of our framework.
The CSM velocity may also be influenced by an accretion disk wind for the case of a compact object companion, as discussed in Section \ref{sec:fastcsm}.

Our binary evolution models are computed assuming that the CSM is ejected from the binary system, but without incorporating the details of the dynamics of that process. Material leaving from the larger lever arm of the L2 point may cause more angular momentum loss than we have realized in our models. As noted in \cite{wu2022}, this may amount to a $\sim 20$ \% increase in the mass transfer rate due to greater orbital contraction. 

Furthermore, we represent the CSM density distribution with a one-dimensional profile, where the mass lost via non-conservative mass transfer and from stellar winds are treated as independent processes. In reality, these two mass loss mechanisms can influence each other; moreover, the magnitude of the stellar wind mass-loss is still uncertain for these stripped stars \citep[e.g.,][]{Tramper16,Yoon17,Sander20,Gotberg23}. 

The uncertainty in stellar wind affects the early radio light curves if the dense CSM is detached, as well as for asymmetric CSM viewed from polar regions. Recent comparison with early light curves of SN Ibc favors larger mass loss rates of $\dot{M}_{\rm w} \gtrsim 10^{-6}~M_\odot\ {\rm yr}^{-1}$ (for wind velocities of $1000$ km s$^{-1}$) during the period shortly before core-collapse \citep{Moriya22}. Such elevated wind mass loss rates are also favored by the observed early-time radio data: the stellar wind could be up to an order of magnitude denser than we currently assume (Equation \ref{eq:Mdotwind}), which would amplify and prolong the early peak generated by interaction between the SN shock and stellar wind. In particular, the early bumps in the light curve visible in Figures \ref{fig:compare_v_csm} and \ref{fig:viewing_angles} would become brighter and peak later, improving the quantitative agreement between our models and observations.

In addition, the parameter space traversed by the 3 GHz light curves in Figure \ref{fig:allmodels_latetime} is occupied by the brightest Type Ib/c SNe in the radio, as seen in light curves at 4--10 GHz within 1 yr from the SN \citep{bietenholz2021}. In contrast, the majority of Type Ib/c SNe peak at around $L_{\nu} \sim 10^{25}$--$10^{27}~\mathrm{erg}\, \mathrm{s}^{-1}\, \mathrm{Hz}^{-1}$. Only a small fraction of all Type Ib/c progenitors may be well-described by our models of non-conservative mass transfer from stripped stars in close binaries, as many other Type Ib/c SN progenitors will have wider binary separations \citep{moriya2015b,ko2025} or will be too massive to expand past their Roche lobes \citep{laplace2020}. However, this tension may also be alleviated by assuming a faster CSM velocity, as discussed in Section \ref{sec:fastcsm}.

Finally, for the asymmetric case of $f_\Omega<1$, we have shown that different viewing angles of the system reveal variations in the early-time radio light curve due to interaction with the stellar wind, which contributes to the light curve on timescales of weeks for a face-on viewing angle. The emission at timescales of years to decades will not strongly depend on the line of sight, as then the CSM is optically thin to GHz emission at all viewing angles. Yet our predictions at earlier phases remain uncertain due to the viewing-angle dependence of free-free absorption. We encourage future work on developing more sophisticated viewing-angle dependent light curves when modeling both the early and late-time radio signals by CSM from binary interaction.

\section{Summary and Conclusions}
\label{sec:conclusions}
In this work, we calculate light curves of radio emission produced by the interaction between SN ejecta and a distribution of dense CSM.
To generate this dense CSM, we model the mass transfer history of binary systems consisting of a stripped star with a low-mass compact companion; these systems represent likely progenitors of hydrogen-poor CCSNe. We consider non-conservative mass transfer that forms a circumbinary outflow, which we find is typically lost at rates of $\sim 10^{-4}\, M_{\odot}\, \rm{yr}^{-1}$. Thus, the CSM density profiles of our models are typically orders of magnitude denser than that created by stellar winds, and for lower-mass progenitors can exhibit features such as detached shells. 

We find that the interaction between SN ejecta and dense CSM originating from binary mass transfer can give rise to highly luminous radio emission. 
Our radio light curves at 3 GHz peak at $L_{\nu}\sim 10^{27}$--$10^{29}~\mathrm{erg}\, \mathrm{s}^{-1}\, \mathrm{Hz}^{-1}$ on timescales of years. Different assumptions for the asymmetry of the CSM, as well as reasonable values for the fractions of shock energy density in magnetic fields $\epsilon_B$ and relativistic electrons $\epsilon_E$, all produce bright emission at late times that is comparable to observed radio emission from a sample of H-poor events at a few--$ 20$ yr.

Events with early radio data exhibit early peaks in the radio emission, which favor models that include an added contribution from nearby low-density material. We demonstrate that accelerating a detached CSM shell, as is characteristically produced by low-mass donors, to higher velocities of $\approx 10^3~\mathrm{km}\ \mathrm{s}^{-1}$ can produce light curves with multiple peaks. These comprise emission on timescales of $\sim$weeks due to interaction with the stellar wind, followed by bright late-time re-brightening, as seen in the radio data for several events, once the SN shock reaches the dense CSM generated by mass transfer. For asymmetric CSM, we also examine how alternate viewing angles of the system can unveil the interaction of the SN shock with the stellar wind that populates the polar regions, an effect which similarly produces an early peak.

In future work, it will be important to apply our forward modeling of radio emission towards other potential scenarios to produce different CSM morphologies, such as common envelope evolution and mass transfer in binaries with larger mass ratios $M_c/M_*$. As the framework to model radio emission described in this paper is designed for arbitrary CSM density profiles, it can be employed to study CSM with properties beyond those modeled here. This may extend to CSM from other systems experiencing stable binary mass transfer, including H-rich Type II SN progenitors (e.g., \citealt{matsuoka2024,ercolino2024,soria2025}) and low-mass stripped progenitors of SNe Ibn and USSNe (e.g., \citealt{wu2022}). Finally, our framework for modeling the dynamical shock evolution can also be applied to other late-time probes of dense CSM, such as the infrared \citep[e.g.,][]{myers2024}, X-rays/$\gamma$-rays, and high-energy neutrinos \citep[e.g.,][]{Murase19, Sarmah23}.

\begin{acknowledgments}
    We thank Raphael Baer-Way, Wynn Jacobson-Galan, Wenbin Lu, Raffaella Margutti, Brenna Mockler, Kohta Murase, and Anthony Piro for helpful discussions. S.C.W. is grateful for support from the Carnegie Theoretical Astrophysics Center. D.T. is grateful for support from the Sherman Fairchild Postdoctoral Fellowship at Caltech. 
\end{acknowledgments}

\bibliography{bib}

\begin{thebibliography}{}
\expandafter\ifx\csname natexlab\endcsname\relax\def\natexlab#1{#1}\fi
\providecommand{\url}[1]{\href{#1}{#1}}
\providecommand{\dodoi}[1]{doi:~\href{http://doi.org/#1}{\nolinkurl{#1}}}
\providecommand{\doeprint}[1]{\href{http://ascl.net/#1}{\nolinkurl{http://ascl.net/#1}}}
\providecommand{\doarXiv}[1]{\href{https://arxiv.org/abs/#1}{\nolinkurl{https://arxiv.org/abs/#1}}}

\bibitem[{{Anderson} {et~al.}(2017){Anderson}, {Horesh}, {Mooley}, {Rushton},
  {Fender}, {Staley}, {Argo}, {Beswick}, {Hancock}, {P{\'e}rez-Torres},
  {Perrott}, {Plotkin}, {Pretorius}, {Rumsey}, \&
  {Titterington}}]{anderson2017}
{Anderson}, G.~E., {Horesh}, A., {Mooley}, K.~P., {et~al.} 2017, \mnras, 466,
  3648, \dodoi{10.1093/mnras/stw3310}

\bibitem[{{Arnett} \& {Meakin}(2011)}]{Arnett2011}
{Arnett}, W.~D., \& {Meakin}, C. 2011, \apj, 733, 78,
  \dodoi{10.1088/0004-637X/733/2/78}

\bibitem[{{Bietenholz} {et~al.}(2021){Bietenholz}, {Bartel}, {Argo}, {Dua},
  {Ryder}, \& {Soderberg}}]{bietenholz2021}
{Bietenholz}, M.~F., {Bartel}, N., {Argo}, M., {et~al.} 2021, \apj, 908, 75,
  \dodoi{10.3847/1538-4357/abccd9}

\bibitem[{{Brethauer} {et~al.}(2022){Brethauer}, {Margutti}, {Milisavljevic},
  {Bietenholz}, {Chornock}, {Coppejans}, {De Colle}, {Hajela}, {Terreran},
  {Vargas}, {DeMarchi}, {Harris}, {Jacobson-Gal{\'a}n}, {Kamble}, {Patnaude},
  \& {Stroh}}]{Brethauer22}
{Brethauer}, D., {Margutti}, R., {Milisavljevic}, D., {et~al.} 2022, \apj, 939,
  105, \dodoi{10.3847/1538-4357/ac8b14}

\bibitem[{{Bruch} {et~al.}(2021){Bruch}, {Gal-Yam}, {Schulze}, {Yaron}, {Yang},
  {Soumagnac}, {Rigault}, {Strotjohann}, {Ofek}, {Sollerman}, {Masci},
  {Barbarino}, {Ho}, {Fremling}, {Perley}, {Nordin}, {Cenko}, {Adams},
  {Adreoni}, {Bellm}, {Blagorodnova}, {Bulla}, {Burdge}, {De}, {Dhawan},
  {Drake}, {Duev}, {Dugas}, {Graham}, {Graham}, {Irani}, {Jencson},
  {Karamehmetoglu}, {Kasliwal}, {Kim}, {Kulkarni}, {Kupfer}, {Liang},
  {Mahabal}, {Miller}, {Prince}, {Riddle}, {Sharma}, {Smith}, {Taddia},
  {Taggart}, {Walters}, \& {Yan}}]{bruch2021}
{Bruch}, R.~J., {Gal-Yam}, A., {Schulze}, S., {et~al.} 2021, \apj, 912, 46,
  \dodoi{10.3847/1538-4357/abef05}

\bibitem[{{Bruch} {et~al.}(2023){Bruch}, {Gal-Yam}, {Yaron}, {Chen},
  {Strotjohann}, {Irani}, {Zimmerman}, {Schulze}, {Yang}, {Kim}, {Bulla},
  {Sollerman}, {Rigault}, {Ofek}, {Soumagnac}, {Masci}, {Fremling}, {Perley},
  {Nordin}, {Cenko}, {Ho}, {Adams}, {Adreoni}, {Bellm}, {Blagorodnova},
  {Burdge}, {De}, {Dekany}, {Dhawan}, {Drake}, {Duev}, {Graham}, {Graham},
  {Jencson}, {Karamehmetoglu}, {Kasliwal}, {Kulkarni}, {Miller}, {Neill},
  {Prince}, {Riddle}, {Rusholme}, {Sharma}, {Smith}, {Sravan}, {Taggart},
  {Walters}, \& {Yan}}]{bruch2023}
{Bruch}, R.~J., {Gal-Yam}, A., {Yaron}, O., {et~al.} 2023, \apj, 952, 119,
  \dodoi{10.3847/1538-4357/acd8be}

\bibitem[{{Chevalier}(1998)}]{Chevalier98}
{Chevalier}, R.~A. 1998, \apj, 499, 810, \dodoi{10.1086/305676}

\bibitem[{{Chevalier}(2012)}]{chevalier12_CE}
---. 2012, \apjl, 752, L2, \dodoi{10.1088/2041-8205/752/1/L2}

\bibitem[{{Chevalier} \& {Fransson}(2006)}]{Chevalier06}
{Chevalier}, R.~A., \& {Fransson}, C. 2006, \apj, 651, 381,
  \dodoi{10.1086/507606}

\bibitem[{{Chevalier} \& {Soker}(1989)}]{Chevalier89}
{Chevalier}, R.~A., \& {Soker}, N. 1989, \apj, 341, 867, \dodoi{10.1086/167545}

\bibitem[{{Crowther}(2007)}]{crowther2007}
{Crowther}, P.~A. 2007, \araa, 45, 177,
  \dodoi{10.1146/annurev.astro.45.051806.110615}

\bibitem[{{DeMarchi} {et~al.}(2022){DeMarchi}, {Margutti}, {Dittman},
  {Brunthaler}, {Milisavljevic}, {Bietenholz}, {Stauffer}, {Brethauer},
  {Coppejans}, {Auchettl}, {Alexander}, {Kilpatrick}, {Bright}, {Kelley},
  {Stroh}, \& {Jacobson-Gal{\'a}n}}]{demarchi2022}
{DeMarchi}, L., {Margutti}, R., {Dittman}, J., {et~al.} 2022, \apj, 938, 84,
  \dodoi{10.3847/1538-4357/ac8c26}

\bibitem[{{Dessart} {et~al.}(2022){Dessart}, {Hillier}, \&
  {Kuncarayakti}}]{dessart2022}
{Dessart}, L., {Hillier}, D.~J., \& {Kuncarayakti}, H. 2022, \aap, 658, A130,
  \dodoi{10.1051/0004-6361/202142436}

\bibitem[{{Dessart} {et~al.}(2015){Dessart}, {Hillier}, {Woosley}, {Livne},
  {Waldman}, {Yoon}, \& {Langer}}]{Dessart15}
{Dessart}, L., {Hillier}, D.~J., {Woosley}, S., {et~al.} 2015, \mnras, 453,
  2189, \dodoi{10.1093/mnras/stv1747}

\bibitem[{{Dewi} \& {Pols}(2003)}]{dewi2003}
{Dewi}, J.~D.~M., \& {Pols}, O.~R. 2003, \mnras, 344, 629,
  \dodoi{10.1046/j.1365-8711.2003.06844.x}

\bibitem[{{Dong} {et~al.}(2021){Dong}, {Hallinan}, {Nakar}, {Ho}, {Hughes},
  {Hotokezaka}, {Myers}, {De}, {Mooley}, {Ravi}, {Horesh}, {Kasliwal}, \&
  {Kulkarni}}]{Dong21}
{Dong}, D.~Z., {Hallinan}, G., {Nakar}, E., {et~al.} 2021, Science, 373, 1125,
  \dodoi{10.1126/science.abg6037}

\bibitem[{{Dong} {et~al.}(2024){Dong}, {Tsuna}, {Valenti}, {Sand}, {Andrews},
  {Bostroem}, {Hosseinzadeh}, {Hoang}, {Jha}, {Janzen}, {Jencson}, {Lundquist},
  {Mehta}, {Ravi}, {Meza Retamal}, {Pearson}, {Shrestha}, {Bonanos}, {Howell},
  {Smith}, {Farah}, {Hiramatsu}, {Itagaki}, {McCully}, {Newsome}, {Padilla
  Gonzalez}, {Paraskeva}, {Pellegrino}, {Terreran}, {Haislip}, {Kouprianov}, \&
  {Reichart}}]{dong2024}
{Dong}, Y., {Tsuna}, D., {Valenti}, S., {et~al.} 2024, \apj, 977, 254,
  \dodoi{10.3847/1538-4357/ad8de6}

\bibitem[{{Drout} {et~al.}(2011){Drout}, {Soderberg}, {Gal-Yam}, {Cenko},
  {Fox}, {Leonard}, {Sand}, {Moon}, {Arcavi}, \& {Green}}]{Drout11}
{Drout}, M.~R., {Soderberg}, A.~M., {Gal-Yam}, A., {et~al.} 2011, \apj, 741,
  97, \dodoi{10.1088/0004-637X/741/2/97}

\bibitem[{{Ercolino} {et~al.}(2024){Ercolino}, {Jin}, {Langer}, \&
  {Dessart}}]{ercolino2024}
{Ercolino}, A., {Jin}, H., {Langer}, N., \& {Dessart}, L. 2024, \aap, 685, A58,
  \dodoi{10.1051/0004-6361/202347646}

\bibitem[{{F{\"o}rster} {et~al.}(2018){F{\"o}rster}, {Moriya}, {Maureira},
  {Anderson}, {Blinnikov}, {Bufano}, {Cabrera-Vives}, {Clocchiatti}, {de
  Jaeger}, {Est{\'e}vez}, {Galbany}, {Gonz{\'a}lez-Gait{\'a}n}, {Gr{\"a}fener},
  {Hamuy}, {Hsiao}, {Huentelemu}, {Huijse}, {Kuncarayakti}, {Mart{\'\i}nez},
  {Medina}, {Olivares E.}, {Pignata}, {Razza}, {Reyes}, {San Mart{\'\i}n},
  {Smith}, {Vera}, {Vivas}, {de Ugarte Postigo}, {Yoon}, {Ashall}, {Fraser},
  {Gal-Yam}, {Kankare}, {Le Guillou}, {Mazzali}, {Walton}, \&
  {Young}}]{forster2018}
{F{\"o}rster}, F., {Moriya}, T.~J., {Maureira}, J.~C., {et~al.} 2018, Nature
  Astronomy, 2, 808, \dodoi{10.1038/s41550-018-0563-4}

\bibitem[{{Fouka} \& {Ouichaoui}(2013)}]{Fouka2013}
{Fouka}, M., \& {Ouichaoui}, S. 2013, Research in Astronomy and Astrophysics,
  13, 680, \dodoi{10.1088/1674-4527/13/6/007}

\bibitem[{{Fuller}(2017)}]{Fuller2017}
{Fuller}, J. 2017, \mnras, 470, 1642, \dodoi{10.1093/mnras/stx1314}

\bibitem[{{Fuller} \& {Ro}(2018)}]{Fuller2018}
{Fuller}, J., \& {Ro}, S. 2018, \mnras, 476, 1853, \dodoi{10.1093/mnras/sty369}

\bibitem[{{Gal-Yam} {et~al.}(2022){Gal-Yam}, {Bruch}, {Schulze}, {Yang},
  {Perley}, {Irani}, {Sollerman}, {Kool}, {Soumagnac}, {Yaron}, {Strotjohann},
  {Zimmerman}, {Barbarino}, {Kulkarni}, {Kasliwal}, {De}, {Yao}, {Fremling},
  {Yan}, {Ofek}, {Fransson}, {Filippenko}, {Zheng}, {Brink}, {Copperwheat},
  {Foley}, {Brown}, {Siebert}, {Leloudas}, {Cabrera-Lavers}, {Garcia-Alvarez},
  {Marante-Barreto}, {Frederick}, {Hung}, {Wheeler}, {Vink{\'o}}, {Thomas},
  {Graham}, {Duev}, {Drake}, {Dekany}, {Bellm}, {Rusholme}, {Shupe},
  {Andreoni}, {Sharma}, {Riddle}, {van Roestel}, \& {Knezevic}}]{Galyam22}
{Gal-Yam}, A., {Bruch}, R., {Schulze}, S., {et~al.} 2022, \nat, 601, 201,
  \dodoi{10.1038/s41586-021-04155-1}

\bibitem[{{Ganss} {et~al.}(2022){Ganss}, {Pledger}, {Sansom}, {James}, {Puls},
  \& {Habergham-Mawson}}]{Ganss22}
{Ganss}, R., {Pledger}, J.~L., {Sansom}, A.~E., {et~al.} 2022, \mnras, 512,
  1541, \dodoi{10.1093/mnras/stac625}

\bibitem[{Gilkis {et~al.}(2025)Gilkis, Laplace, Arcavi, Shenar, \&
  Schneider}]{Gilkis2025}
Gilkis, A., Laplace, E., Arcavi, I., Shenar, T., \& Schneider, F. R.~N. 2025,
  Monthly Notices of the Royal Astronomical Society, 540, 3094,
  \dodoi{10.1093/mnras/staf884}

\bibitem[{{G{\"o}tberg} {et~al.}(2017){G{\"o}tberg}, {de Mink}, \&
  {Groh}}]{Gotberg17}
{G{\"o}tberg}, Y., {de Mink}, S.~E., \& {Groh}, J.~H. 2017, \aap, 608, A11,
  \dodoi{10.1051/0004-6361/201730472}

\bibitem[{{G{\"o}tberg} {et~al.}(2023){G{\"o}tberg}, {Drout}, {Ji}, {Groh},
  {Ludwig}, {Crowther}, {Smith}, {de Koter}, \& {de Mink}}]{Gotberg23}
{G{\"o}tberg}, Y., {Drout}, M.~R., {Ji}, A.~P., {et~al.} 2023, \apj, 959, 125,
  \dodoi{10.3847/1538-4357/ace5a3}

\bibitem[{{Granot} {et~al.}(2018){Granot}, {De Colle}, \&
  {Ramirez-Ruiz}}]{granot2018}
{Granot}, J., {De Colle}, F., \& {Ramirez-Ruiz}, E. 2018, \mnras, 481, 2711,
  \dodoi{10.1093/mnras/sty2454}

\bibitem[{{Habets}(1986)}]{habets1986}
{Habets}, G.~M.~H.~J. 1986, \aap, 167, 61

\bibitem[{{Hosseinzadeh} {et~al.}(2017){Hosseinzadeh}, {Arcavi}, {Valenti},
  {McCully}, {Howell}, {Johansson}, {Sollerman}, {Pastorello}, {Benetti},
  {Cao}, {Cenko}, {Clubb}, {Corsi}, {Duggan}, {Elias-Rosa}, {Filippenko},
  {Fox}, {Fremling}, {Horesh}, {Karamehmetoglu}, {Kasliwal}, {Marion}, {Ofek},
  {Sand}, {Taddia}, {Zheng}, {Fraser}, {Gal-Yam}, {Inserra}, {Laher}, {Masci},
  {Rebbapragada}, {Smartt}, {Smith}, {Sullivan}, {Surace}, \&
  {Wo{\'z}niak}}]{hosseinzadeh2017}
{Hosseinzadeh}, G., {Arcavi}, I., {Valenti}, S., {et~al.} 2017, \apj, 836, 158,
  \dodoi{10.3847/1538-4357/836/2/158}

\bibitem[{{Jacobson-Gal{\'a}n} {et~al.}(2024){Jacobson-Gal{\'a}n}, {Dessart},
  {Davis}, {Kilpatrick}, {Margutti}, {Foley}, {Chornock}, {Terreran},
  {Hiramatsu}, {Newsome}, {Padilla Gonzalez}, {Pellegrino}, {Howell},
  {Filippenko}, {Anderson}, {Angus}, {Auchettl}, {Bostroem}, {Brink},
  {Cartier}, {Coulter}, {de Boer}, {Drout}, {Earl}, {Ertini}, {Farah},
  {Farias}, {Gall}, {Gao}, {Gerlach}, {Guo}, {Haynie}, {Hosseinzadeh}, {Ibik},
  {Jha}, {Jones}, {Langeroodi}, {LeBaron}, {Magnier}, {Piro}, {Raimundo},
  {Rest}, {Rest}, {Rich}, {Rojas-Bravo}, {Sears}, {Taggart}, {Villar},
  {Wainscoat}, {Wang}, {Wasserman}, {Yan}, {Yang}, {Zhang}, \&
  {Zheng}}]{jacobsongalan2024}
{Jacobson-Gal{\'a}n}, W.~V., {Dessart}, L., {Davis}, K.~W., {et~al.} 2024,
  \apj, 970, 189, \dodoi{10.3847/1538-4357/ad4a2a}

\bibitem[{{Kamble} {et~al.}(2014){Kamble}, {Soderberg}, {Chomiuk}, {Margutti},
  {Medvedev}, {Milisavljevic}, {Chakraborti}, {Chevalier}, {Chugai},
  {Dittmann}, {Drout}, {Fransson}, {Nakar}, \& {Sanders}}]{kamble2014}
{Kamble}, A., {Soderberg}, A.~M., {Chomiuk}, L., {et~al.} 2014, \apj, 797, 2,
  \dodoi{10.1088/0004-637X/797/1/2}

\bibitem[{{Ko} {et~al.}(2025){Ko}, {Kinugawa}, {Tsuna}, {Hirai}, \&
  {Takei}}]{ko2025}
{Ko}, T., {Kinugawa}, T., {Tsuna}, D., {Hirai}, R., \& {Takei}, Y. 2025, arXiv
  e-prints, arXiv:2506.00931, \dodoi{10.48550/arXiv.2506.00931}

\bibitem[{{Kolb} \& {Ritter}(1990)}]{kolb1990}
{Kolb}, U., \& {Ritter}, H. 1990, \aap, 236, 385

\bibitem[{{Laplace} {et~al.}(2020){Laplace}, {G{\"o}tberg}, {de Mink},
  {Justham}, \& {Farmer}}]{laplace2020}
{Laplace}, E., {G{\"o}tberg}, Y., {de Mink}, S.~E., {Justham}, S., \& {Farmer},
  R. 2020, \aap, 637, A6, \dodoi{10.1051/0004-6361/201937300}

\bibitem[{{Lau} {et~al.}(2022{\natexlab{a}}){Lau}, {Hirai},
  {Gonz{\'a}lez-Bol{\'\i}var}, {Price}, {De Marco}, \& {Mandel}}]{lau2022a}
{Lau}, M. Y.~M., {Hirai}, R., {Gonz{\'a}lez-Bol{\'\i}var}, M., {et~al.}
  2022{\natexlab{a}}, \mnras, 512, 5462, \dodoi{10.1093/mnras/stac049}

\bibitem[{{Lau} {et~al.}(2024){Lau}, {Hirai}, {Mandel}, \& {Tout}}]{Lau24}
{Lau}, M. Y.~M., {Hirai}, R., {Mandel}, I., \& {Tout}, C.~A. 2024, \apjl, 966,
  L7, \dodoi{10.3847/2041-8213/ad3d50}

\bibitem[{{Lau} {et~al.}(2022{\natexlab{b}}){Lau}, {Hirai}, {Price}, \&
  {Mandel}}]{lau2022b}
{Lau}, M. Y.~M., {Hirai}, R., {Price}, D.~J., \& {Mandel}, I.
  2022{\natexlab{b}}, \mnras, 516, 4669, \dodoi{10.1093/mnras/stac2490}

\bibitem[{{Lau} {et~al.}(2025){Lau}, {Hirai}, {Price}, {Mandel}, \&
  {Bate}}]{Lau25}
{Lau}, M. Y.~M., {Hirai}, R., {Price}, D.~J., {Mandel}, I., \& {Bate}, M.~R.
  2025, arXiv e-prints, arXiv:2503.20506, \dodoi{10.48550/arXiv.2503.20506}

\bibitem[{{Levinson} \& {Nakar}(2020)}]{Levinson20}
{Levinson}, A., \& {Nakar}, E. 2020, \physrep, 866, 1,
  \dodoi{10.1016/j.physrep.2020.04.003}

\bibitem[{{Lu} {et~al.}(2023){Lu}, {Fuller}, {Quataert}, \&
  {Bonnerot}}]{lu2022}
{Lu}, W., {Fuller}, J., {Quataert}, E., \& {Bonnerot}, C. 2023, \mnras, 519,
  1409, \dodoi{10.1093/mnras/stac3621}

\bibitem[{{Lyman} {et~al.}(2016){Lyman}, {Bersier}, {James}, {Mazzali},
  {Eldridge}, {Fraser}, \& {Pian}}]{Lyman16}
{Lyman}, J.~D., {Bersier}, D., {James}, P.~A., {et~al.} 2016, \mnras, 457, 328,
  \dodoi{10.1093/mnras/stv2983}

\bibitem[{{MacLeod} {et~al.}(2018){MacLeod}, {Ostriker}, \&
  {Stone}}]{MacLeod2018}
{MacLeod}, M., {Ostriker}, E.~C., \& {Stone}, J.~M. 2018, \apj, 868, 136,
  \dodoi{10.3847/1538-4357/aae9eb}

\bibitem[{{Maeda}(2012)}]{Maeda12}
{Maeda}, K. 2012, \apj, 758, 81, \dodoi{10.1088/0004-637X/758/2/81}

\bibitem[{{Maeda} \& {Moriya}(2022)}]{Maeda22}
{Maeda}, K., \& {Moriya}, T.~J. 2022, \apj, 927, 25,
  \dodoi{10.3847/1538-4357/ac4672}

\bibitem[{{Marchant}(2025)}]{marchant2025}
{Marchant}, P. 2025, arXiv e-prints, arXiv:2503.16099,
  \dodoi{10.48550/arXiv.2503.16099}

\bibitem[{{Marchant} {et~al.}(2021){Marchant}, {Pappas}, {Gallegos-Garcia},
  {Berry}, {Taam}, {Kalogera}, \& {Podsiadlowski}}]{marchant2021}
{Marchant}, P., {Pappas}, K. M.~W., {Gallegos-Garcia}, M., {et~al.} 2021, \aap,
  650, A107, \dodoi{10.1051/0004-6361/202039992}

\bibitem[{{Margalit} \& {Quataert}(2024)}]{Margalit24}
{Margalit}, B., \& {Quataert}, E. 2024, \apj, 977, 134,
  \dodoi{10.3847/1538-4357/ad8b47}

\bibitem[{{Margalit} {et~al.}(2022){Margalit}, {Quataert}, \&
  {Ho}}]{Margalit22}
{Margalit}, B., {Quataert}, E., \& {Ho}, A. Y.~Q. 2022, \apj, 928, 122,
  \dodoi{10.3847/1538-4357/ac53b0}

\bibitem[{{Margutti} {et~al.}(2017){Margutti}, {Kamble}, {Milisavljevic},
  {Zapartas}, {de Mink}, {Drout}, {Chornock}, {Risaliti}, {Zauderer},
  {Bietenholz}, {Cantiello}, {Chakraborti}, {Chomiuk}, {Fong}, {Grefenstette},
  {Guidorzi}, {Kirshner}, {Parrent}, {Patnaude}, {Soderberg}, {Gehrels}, \&
  {Harrison}}]{Margutti17}
{Margutti}, R., {Kamble}, A., {Milisavljevic}, D., {et~al.} 2017, \apj, 835,
  140, \dodoi{10.3847/1538-4357/835/2/140}

\bibitem[{{Matsuoka} \& {Sawada}(2024)}]{matsuoka2024}
{Matsuoka}, T., \& {Sawada}, R. 2024, \apj, 963, 105,
  \dodoi{10.3847/1538-4357/ad1829}

\bibitem[{{Matzner} \& {McKee}(1999)}]{Matzner99}
{Matzner}, C.~D., \& {McKee}, C.~F. 1999, \apj, 510, 379,
  \dodoi{10.1086/306571}

\bibitem[{{Meakin} \& {Arnett}(2006)}]{Meakin2006}
{Meakin}, C.~A., \& {Arnett}, D. 2006, \apjl, 637, L53, \dodoi{10.1086/500544}

\bibitem[{{Meakin} \& {Arnett}(2007)}]{Meakin2007a}
---. 2007, \apj, 665, 690, \dodoi{10.1086/519372}

\bibitem[{{Mezger} \& {Henderson}(1967)}]{Mezger67}
{Mezger}, P.~G., \& {Henderson}, A.~P. 1967, \apj, 147, 471,
  \dodoi{10.1086/149030}

\bibitem[{{Moriya} \& {Langer}(2015)}]{moriya2015}
{Moriya}, T.~J., \& {Langer}, N. 2015, \aap, 573, A18,
  \dodoi{10.1051/0004-6361/201424957}

\bibitem[{{Moriya} {et~al.}(2015){Moriya}, {Liu}, \& {Izzard}}]{moriya2015b}
{Moriya}, T.~J., {Liu}, Z.-W., \& {Izzard}, R.~G. 2015, \mnras, 450, 3264,
  \dodoi{10.1093/mnras/stv934}

\bibitem[{{Moriya} {et~al.}(2013){Moriya}, {Maeda}, {Taddia}, {Sollerman},
  {Blinnikov}, \& {Sorokina}}]{Moriya_et_al_13}
{Moriya}, T.~J., {Maeda}, K., {Taddia}, F., {et~al.} 2013, \mnras, 435, 1520,
  \dodoi{10.1093/mnras/stt1392}

\bibitem[{{Moriya} \& {Yoon}(2022)}]{Moriya22}
{Moriya}, T.~J., \& {Yoon}, S.-C. 2022, \mnras, 513, 5606,
  \dodoi{10.1093/mnras/stac1271}

\bibitem[{{Morozova} {et~al.}(2017){Morozova}, {Piro}, \&
  {Valenti}}]{Morozova2017}
{Morozova}, V., {Piro}, A.~L., \& {Valenti}, S. 2017, \apj, 838, 28,
  \dodoi{10.3847/1538-4357/aa6251}

\bibitem[{{Murase}(2018)}]{Murase18}
{Murase}, K. 2018, \prd, 97, 081301, \dodoi{10.1103/PhysRevD.97.081301}

\bibitem[{{Murase}(2024)}]{Murase24}
---. 2024, \prd, 109, 103020, \dodoi{10.1103/PhysRevD.109.103020}

\bibitem[{{Murase} {et~al.}(2019){Murase}, {Franckowiak}, {Maeda}, {Margutti},
  \& {Beacom}}]{Murase19}
{Murase}, K., {Franckowiak}, A., {Maeda}, K., {Margutti}, R., \& {Beacom},
  J.~F. 2019, \apj, 874, 80, \dodoi{10.3847/1538-4357/ab0422}

\bibitem[{{Myers} {et~al.}(2024){Myers}, {De}, {Yan}, {Jencson}, {Earley},
  {Fremling}, {Hiramatsu}, {Kasliwal}, {Lau}, {MacLeod}, {Masterson},
  {Panagiotou}, {Simcoe}, \& {Tinyanont}}]{myers2024}
{Myers}, C., {De}, K., {Yan}, L., {et~al.} 2024, \apj, 976, 230,
  \dodoi{10.3847/1538-4357/ad8922}

\bibitem[{{Nugis} \& {Lamers}(2000)}]{Nugis2000}
{Nugis}, T., \& {Lamers}, H.~J.~G.~L.~M. 2000, \aap, 360, 227

\bibitem[{{Orlando} {et~al.}(2024){Orlando}, {Greco}, {Hirai}, {Matsuoka},
  {Miceli}, {Nagataki}, {Ono}, {Chen}, {Milisavljevic}, {Patnaude}, {Bocchino},
  \& {Elias-Rosa}}]{orlando2024}
{Orlando}, S., {Greco}, E., {Hirai}, R., {et~al.} 2024, \apj, 977, 118,
  \dodoi{10.3847/1538-4357/ad8ac8}

\bibitem[{{Pastorello} {et~al.}(2008){Pastorello}, {Mattila}, {Zampieri},
  {Della Valle}, {Smartt}, {Valenti}, {Agnoletto}, {Benetti}, {Benn}, {Branch},
  {Cappellaro}, {Dennefeld}, {Eldridge}, {Gal-Yam}, {Harutyunyan}, {Hunter},
  {Kjeldsen}, {Lipkin}, {Mazzali}, {Milne}, {Navasardyan}, {Ofek}, {Pian},
  {Shemmer}, {Spiro}, {Stathakis}, {Taubenberger}, {Turatto}, \&
  {Yamaoka}}]{Pastorello08}
{Pastorello}, A., {Mattila}, S., {Zampieri}, L., {et~al.} 2008, \mnras, 389,
  113, \dodoi{10.1111/j.1365-2966.2008.13602.x}

\bibitem[{{Paxton} {et~al.}(2011){Paxton}, {Bildsten}, {Dotter}, {Herwig},
  {Lesaffre}, \& {Timmes}}]{mesa2011}
{Paxton}, B., {Bildsten}, L., {Dotter}, A., {et~al.} 2011, \apjs, 192, 3,
  \dodoi{10.1088/0067-0049/192/1/3}

\bibitem[{{Paxton} {et~al.}(2013){Paxton}, {Cantiello}, {Arras}, {Bildsten},
  {Brown}, {Dotter}, {Mankovich}, {Montgomery}, {Stello}, {Timmes}, \&
  {Townsend}}]{mesa2013}
{Paxton}, B., {Cantiello}, M., {Arras}, P., {et~al.} 2013, \apjs, 208, 4,
  \dodoi{10.1088/0067-0049/208/1/4}

\bibitem[{{Paxton} {et~al.}(2015){Paxton}, {Marchant}, {Schwab}, {Bauer},
  {Bildsten}, {Cantiello}, {Dessart}, {Farmer}, {Hu}, {Langer}, {Townsend},
  {Townsley}, \& {Timmes}}]{mesa2015}
{Paxton}, B., {Marchant}, P., {Schwab}, J., {et~al.} 2015, \apjs, 220, 15,
  \dodoi{10.1088/0067-0049/220/1/15}

\bibitem[{{Paxton} {et~al.}(2018){Paxton}, {Schwab}, {Bauer}, {Bildsten},
  {Blinnikov}, {Duffell}, {Farmer}, {Goldberg}, {Marchant}, {Sorokina},
  {Thoul}, {Townsend}, \& {Timmes}}]{mesa2018}
{Paxton}, B., {Schwab}, J., {Bauer}, E.~B., {et~al.} 2018, \apjs, 234, 34,
  \dodoi{10.3847/1538-4365/aaa5a8}

\bibitem[{{Paxton} {et~al.}(2019){Paxton}, {Smolec}, {Schwab}, {Gautschy},
  {Bildsten}, {Cantiello}, {Dotter}, {Farmer}, {Goldberg}, {Jermyn}, {Kanbur},
  {Marchant}, {Thoul}, {Townsend}, {Wolf}, {Zhang}, \& {Timmes}}]{mesa2019}
{Paxton}, B., {Smolec}, R., {Schwab}, J., {et~al.} 2019, \apjs, 243, 10,
  \dodoi{10.3847/1538-4365/ab2241}

\bibitem[{{Pejcha} {et~al.}(2016){Pejcha}, {Metzger}, \& {Tomida}}]{pejcha2016}
{Pejcha}, O., {Metzger}, B.~D., \& {Tomida}, K. 2016, \mnras, 455, 4351,
  \dodoi{10.1093/mnras/stv2592}

\bibitem[{{Pellegrino} {et~al.}(2022){Pellegrino}, {Howell}, {Vink{\'o}},
  {Gangopadhyay}, {Xiang}, {Arcavi}, {Brown}, {Burke}, {Hiramatsu},
  {Hosseinzadeh}, {Li}, {McCully}, {Misra}, {Newsome}, {Gonzalez}, {Pritchard},
  {Valenti}, {Wang}, \& {Zhang}}]{pellegrino2022}
{Pellegrino}, C., {Howell}, D.~A., {Vink{\'o}}, J., {et~al.} 2022, \apj, 926,
  125, \dodoi{10.3847/1538-4357/ac3e63}

\bibitem[{{Pooley} {et~al.}(2019){Pooley}, {Wheeler}, {Vink{\'o}}, {Dwarkadas},
  {Szalai}, {Silverman}, {Griesel}, {McCullough}, {Marion}, \&
  {MacQueen}}]{pooley2019}
{Pooley}, D., {Wheeler}, J.~C., {Vink{\'o}}, J., {et~al.} 2019, \apj, 883, 120,
  \dodoi{10.3847/1538-4357/ab3e36}

\bibitem[{{Quataert} \& {Shiode}(2012)}]{Quataert2012}
{Quataert}, E., \& {Shiode}, J. 2012, \mnras, 423, L92,
  \dodoi{10.1111/j.1745-3933.2012.01264.x}

\bibitem[{{Rocha} {et~al.}(2024){Rocha}, {Kalogera}, {Doctor}, {Andrews},
  {Sun}, {Gossage}, {Bavera}, {Fragos}, {Kovlakas}, {Kruckow}, {Misra},
  {Srivastava}, {Xing}, \& {Zapartas}}]{rocha2024}
{Rocha}, K.~A., {Kalogera}, V., {Doctor}, Z., {et~al.} 2024, \apj, 971, 133,
  \dodoi{10.3847/1538-4357/ad5955}

\bibitem[{{R{\"o}pke} \& {De Marco}(2023)}]{ropke2023}
{R{\"o}pke}, F.~K., \& {De Marco}, O. 2023, Living Reviews in Computational
  Astrophysics, 9, 2, \dodoi{10.1007/s41115-023-00017-x}

\bibitem[{{Rose} {et~al.}(2024){Rose}, {Horesh}, {Murphy}, {Kaplan}, {Sfaradi},
  {Ryder}, {Aloisi}, {Dobie}, {Driessen}, {Fender}, {Green}, {Leung}, {Lenc},
  {Qiu}, \& {Williams-Baldwin}}]{Rose24}
{Rose}, K., {Horesh}, A., {Murphy}, T., {et~al.} 2024, \mnras, 534, 3853,
  \dodoi{10.1093/mnras/stae2289}

\bibitem[{{Rybicki} \& {Lightman}(1979)}]{Radipro}
{Rybicki}, G.~B., \& {Lightman}, A.~P. 1979, {Radiative processes in
  astrophysics}

\bibitem[{{Sander} \& {Vink}(2020)}]{Sander20}
{Sander}, A. A.~C., \& {Vink}, J.~S. 2020, \mnras, 499, 873,
  \dodoi{10.1093/mnras/staa2712}

\bibitem[{{Sarmah} {et~al.}(2023){Sarmah}, {Chakraborty}, {Tamborra}, \&
  {Auchettl}}]{Sarmah23}
{Sarmah}, P., {Chakraborty}, S., {Tamborra}, I., \& {Auchettl}, K. 2023, \prd,
  108, 103033, \dodoi{10.1103/PhysRevD.108.103033}

\bibitem[{{Schlegel}(1990)}]{Schlegel90}
{Schlegel}, E.~M. 1990, \mnras, 244, 269

\bibitem[{{Smith}(2017)}]{smith2017}
{Smith}, N. 2017, in Handbook of Supernovae, ed. A.~W. {Alsabti} \&
  P.~{Murdin}, 403, \dodoi{10.1007/978-3-319-21846-5_38}

\bibitem[{{Soria} {et~al.}(2025){Soria}, {Russell}, {Wiston}, {Cheng},
  {Margutti}, {Rose}, {Ryder}, \& {Terreran}}]{soria2025}
{Soria}, R., {Russell}, T.~D., {Wiston}, E., {et~al.} 2025, arXiv e-prints,
  arXiv:2502.01740, \dodoi{10.48550/arXiv.2502.01740}

\bibitem[{{Stroh} {et~al.}(2021){Stroh}, {Terreran}, {Coppejans}, {Bright},
  {Margutti}, {Bietenholz}, {De Colle}, {DeMarchi}, {Duran}, {Milisavljevic},
  {Murase}, {Paterson}, \& {Williams}}]{Stroh21}
{Stroh}, M.~C., {Terreran}, G., {Coppejans}, D.~L., {et~al.} 2021, \apjl, 923,
  L24, \dodoi{10.3847/2041-8213/ac375e}

\bibitem[{{Strotjohann} {et~al.}(2021){Strotjohann}, {Ofek}, {Gal-Yam},
  {Bruch}, {Schulze}, {Shaviv}, {Sollerman}, {Filippenko}, {Yaron}, {Fremling},
  {Nordin}, {Kool}, {Perley}, {Ho}, {Yang}, {Yao}, {Soumagnac}, {Graham},
  {Barbarino}, {Tartaglia}, {De}, {Goldstein}, {Cook}, {Brink}, {Taggart},
  {Yan}, {Lunnan}, {Kasliwal}, {Kulkarni}, {Nugent}, {Masci}, {Rosnet},
  {Adams}, {Andreoni}, {Bagdasaryan}, {Bellm}, {Burdge}, {Duev}, {Dugas},
  {Frederick}, {Goldwasser}, {Hankins}, {Irani}, {Karambelkar}, {Kupfer},
  {Liang}, {Neill}, {Porter}, {Riddle}, {Sharma}, {Short}, {Taddia},
  {Tzanidakis}, {van Roestel}, {Walters}, \& {Zhuang}}]{strotjohann2021}
{Strotjohann}, N.~L., {Ofek}, E.~O., {Gal-Yam}, A., {et~al.} 2021, \apj, 907,
  99, \dodoi{10.3847/1538-4357/abd032}

\bibitem[{{Svirski} \& {Nakar}(2014)}]{2014ApJ...788L..14S}
{Svirski}, G., \& {Nakar}, E. 2014, \apjl, 788, L14,
  \dodoi{10.1088/2041-8205/788/1/L14}

\bibitem[{{Taddia} {et~al.}(2018){Taddia}, {Stritzinger}, {Bersten}, {Baron},
  {Burns}, {Contreras}, {Holmbo}, {Hsiao}, {Morrell}, {Phillips}, {Sollerman},
  \& {Suntzeff}}]{Taddia18}
{Taddia}, F., {Stritzinger}, M.~D., {Bersten}, M., {et~al.} 2018, \aap, 609,
  A136, \dodoi{10.1051/0004-6361/201730844}

\bibitem[{{Tauris} \& {van den Heuvel}(2006)}]{tauris2006}
{Tauris}, T.~M., \& {van den Heuvel}, E.~P.~J. 2006, in Compact stellar X-ray
  sources, ed. W.~H.~G. {Lewin} \& M.~{van der Klis}, Vol.~39, 623--665,
  \dodoi{10.48550/arXiv.astro-ph/0303456}

\bibitem[{{Tauris} {et~al.}(2017){Tauris}, {Kramer}, {Freire}, {Wex}, {Janka},
  {Langer}, {Podsiadlowski}, {Bozzo}, {Chaty}, {Kruckow}, {van den Heuvel},
  {Antoniadis}, {Breton}, \& {Champion}}]{tauris2017}
{Tauris}, T.~M., {Kramer}, M., {Freire}, P.~C.~C., {et~al.} 2017, \apj, 846,
  170, \dodoi{10.3847/1538-4357/aa7e89}

\bibitem[{{Terreran} {et~al.}(2019){Terreran}, {Margutti}, {Bersier},
  {Brimacombe}, {Caprioli}, {Challis}, {Chornock}, {Coppejans}, {Dong},
  {Guidorzi}, {Hurley}, {Kirshner}, {Migliori}, {Milisavljevic}, {Palmer},
  {Prieto}, {Tomasella}, {Marchant}, {Pastorello}, {Shappee}, {Stanek},
  {Stritzinger}, {Benetti}, {Chen}, {DeMarchi}, {Elias-Rosa}, {Gall},
  {Harmanen}, \& {Mattila}}]{terreran2019}
{Terreran}, G., {Margutti}, R., {Bersier}, D., {et~al.} 2019, \apj, 883, 147,
  \dodoi{10.3847/1538-4357/ab3e37}

\bibitem[{{Tramper} {et~al.}(2016){Tramper}, {Sana}, \& {de Koter}}]{Tramper16}
{Tramper}, F., {Sana}, H., \& {de Koter}, A. 2016, \apj, 833, 133,
  \dodoi{10.3847/1538-4357/833/2/133}

\bibitem[{{Tsang} {et~al.}(2022){Tsang}, {Kasen}, \& {Bildsten}}]{tsang2022}
{Tsang}, B. T.~H., {Kasen}, D., \& {Bildsten}, L. 2022, \apj, 936, 28,
  \dodoi{10.3847/1538-4357/ac83bc}

\bibitem[{{Tsuna} {et~al.}(2023{\natexlab{a}}){Tsuna}, {Murase}, \&
  {Moriya}}]{Tsuna2023}
{Tsuna}, D., {Murase}, K., \& {Moriya}, T.~J. 2023{\natexlab{a}}, \apj, 952,
  115, \dodoi{10.3847/1538-4357/acdb71}

\bibitem[{{Tsuna} {et~al.}(2023{\natexlab{b}}){Tsuna}, {Takei}, \&
  {Shigeyama}}]{tsuna2023b}
{Tsuna}, D., {Takei}, Y., \& {Shigeyama}, T. 2023{\natexlab{b}}, \apj, 945,
  104, \dodoi{10.3847/1538-4357/acbbc6}

\bibitem[{{Tsuna} {et~al.}(2024){Tsuna}, {Wu}, {Fuller}, {Dong}, \&
  {Piro}}]{Tsuna2024}
{Tsuna}, D., {Wu}, S.~C., {Fuller}, J., {Dong}, Y., \& {Piro}, A.~L. 2024, The
  Open Journal of Astrophysics, 7, 82, \dodoi{10.33232/001c.123897}

\bibitem[{{Wellons} {et~al.}(2012){Wellons}, {Soderberg}, \&
  {Chevalier}}]{wellons2012}
{Wellons}, S., {Soderberg}, A.~M., \& {Chevalier}, R.~A. 2012, \apj, 752, 17,
  \dodoi{10.1088/0004-637X/752/1/17}

\bibitem[{{Woosley} \& {Heger}(2015)}]{Woosley2015}
{Woosley}, S.~E., \& {Heger}, A. 2015, \apj, 810, 34,
  \dodoi{10.1088/0004-637X/810/1/34}

\bibitem[{{Wu} \& {Fuller}(2021)}]{Wu21}
{Wu}, S., \& {Fuller}, J. 2021, \apj, 906, 3, \dodoi{10.3847/1538-4357/abc87c}

\bibitem[{{Wu} \& {Fuller}(2022{\natexlab{a}})}]{wu2022a}
{Wu}, S.~C., \& {Fuller}, J. 2022{\natexlab{a}}, \apj, 930, 119,
  \dodoi{10.3847/1538-4357/ac660c}

\bibitem[{{Wu} \& {Fuller}(2022{\natexlab{b}})}]{wu2022}
---. 2022{\natexlab{b}}, \apjl, 940, L27, \dodoi{10.3847/2041-8213/ac9b3d}

\bibitem[{{Yoon}(2017)}]{Yoon17}
{Yoon}, S.-C. 2017, \mnras, 470, 3970, \dodoi{10.1093/mnras/stx1496}

\bibitem[{{Zapartas} {et~al.}(2017){Zapartas}, {de Mink}, {Van Dyk}, {Fox},
  {Smith}, {Bostroem}, {de Koter}, {Filippenko}, {Izzard}, {Kelly}, {Neijssel},
  {Renzo}, \& {Ryder}}]{Zapartas17}
{Zapartas}, E., {de Mink}, S.~E., {Van Dyk}, S.~D., {et~al.} 2017, \apj, 842,
  125, \dodoi{10.3847/1538-4357/aa7467}

\end{thebibliography}
{}
\bibliographystyle{aasjournal}

\end{document}